\makeatletter
\expandafter\def\csname ver@fontawesome.sty\endcsname{9999/12/31}
\makeatother
\documentclass[manuscript]{copernicus}

\usepackage{tabularx}
\usepackage{array}

\begin{document}
\nolinenumbers

\title{Estimating subsurface water mass changes with ground-based gravimetry}

\Author[1][i.schmidt@fu-berlin.de]{Isabelle Schmidt}{}
\Author[2]{Jackson Ang’ong’a}{}
\Author[3]{Peter Bauer-Gottwein}{}
\Author[4][]{Stephan Costabel}{}
\Author[5][]{Christian Freier}{}
\Author[5,6][]{Bastian Leykauf}{}
\Author[5][]{Tina Lundgren}{}
\Author[2][]{Miguel Martinez-Dorantes}{}
\Author[6][]{Achim Peters}{}
\Author[7][]{Ernst Rasel}{}
\Author[2]{Enrico Vogt}{}
\Author[4][]{Gregor Willkommen}{}
\Author[6,8][]{Vladimir Schkolnik}{}

\affil[1]{Institute of Geological Sciences, Freie Universität Berlin, 12249 Berlin, Germany}
\affil[2]{Qubig GmbH, Grillparzerstr. 6, 81675 Munich, Germany}
 \affil[3]{University of Copenhagen, Department of Geosciences and Natural Resource Management, Rolighedsvej 23, 1958 Frederiksberg C, Denmark}
 \affil[4]{Federal Institute for Geosciences and Natural Resources (BGR), Wilhelmstrasse 26, 13593 Berlin, Germany}
 \affil[5]{Nomad Atomics GmbH, Rudower Chaussee 29 12489, Berlin, Germany}
 \affil[6]{Humboldt Universität zu Berlin, Berlin, Germany}
 \affil[7]{Institute of Quantum Optics, QUEST-Leibniz Research School, Leibniz University Hannover, Hanover, Germany}
 \affil[8]{TOPTICA Photonics SE, Munich, Germany}

\runningtitle{Estimating subsurface water mass changes with ground-based gravimetry}
\runningauthor{Schmidt, Ang’ong’a, Bauer-Gottwein, Costabel, Freier, Leykauf, Lundgren, Martinez-Dorantes, Peters, Rasel, Vogt, Willkommen, and Schkolnik}

\received{}
\pubdiscuss{}
\revised{}
\accepted{}
\published{}

\firstpage{1}

\maketitle
\renewcommand{\baselinestretch}{1.0}\normalsize

\begin{abstract}
Gravitational force is proportional to the mass of an attracting body; therefore, changes in subsurface mass can be detected using gravimetry. In the near subsurface, mass variations are primarily driven by changes in water storage, meaning that gravity measurements provide direct information on variations in water mass. Traditionally, groundwater and soil water dynamics are monitored using point-scale instruments that require direct installation into the subsurface. In contrast, gravimetry is a non-invasive method that integrates signals over a larger spatial extent.
Recent advancements in gravimeter technology have significantly improved measurement accuracy and long-term stability under field conditions, thereby expanding their potential for hydrological applications. This review summarizes recent developments in gravimetric instrumentation and provides essential background on gravimeters used for ground-based measurements and their application in hydrology, aiming to facilitate their broader use. In addition, it presents an overview of studies in hydrology and hydrogeology that have successfully applied gravimetry to quantify subsurface water storage changes, estimate hydrological parameters, and constrain numerical models. A perspective on future advancement in gravimetry and resulting potential applications is given. 
\end{abstract}

\introduction
Groundwater sustains human livelihoods worldwide by providing water for drinking, sanitation, industry, and irrigation. Recently, it has also gained importance as a medium for sustainable heating and cooling through heat exchange \citep{UNESCO2022}. Stored in vast underground aquifers, groundwater buffers droughts and is typically of high quality, thanks to the natural filtration and degradation processes of the subsurface ecosystem. But groundwater resources are at risk. Due to overexploitation, climate change \citep{Scanlon2023}, and pollution, its quality and quantity are reduced \citep{Basu2022,Schroeter2025}. Addressing this emerging groundwater crisis requires strategic water management to protect this valuable resource. 

Groundwater is stored in different formations, depending on the conditions of the underground. In alluvial aquifers, it is stored in and flows through the pores between the grains \citep{Freeze1979}. In hard rock aquifers, it is stored in the fissures of the weathered rock and, if it exists, the porous matrix \citep{Maloszewski1985}. Conditions may be homogeneous on the small scale, but can become heterogeneous on the larger scale, due to the existence of joints or faults in the lateral extension, and in the vertical direction, because weathering is reduced with depth \citep{Lachassagne2021} (Appendix \ref{sec:hardrock}). In karstic aquifers, the conditions are highly heterogeneous, as water flows through conduits, which can store large water masses during extreme rainfall events \citep{Pivetta2021}. Before reaching the groundwater table of the aquifer (saturated zone, SZ), rainwater infiltrates through the unsaturated zone (UZ) or vadose zone \citep{Freeze1979}. In the SZ, groundwater is stored in unconfined (i.e., the aquifer has a connection to the atmosphere through the vadose zone) and confined (i.e., the aquifer is overlain by a low-permeability layer) conditions \citep{SharpJr2023}.

To effectively manage groundwater, information about the underground and the groundwater body itself is required. For example, Specific yield ($Sy$) represents the dimensionless aquifer storage coefﬁcient in unconfined conditions. It is the volume of water that can be released from the aquifer per change in hydraulic head \citep{Freeze1979, Kennedy2023} and is an essential parameter for groundwater management. Furthermore, water management is often supported by physical modelling of the flow processes in the SZ or coupled modeling of the earth surface and the connection with the SZ (unsaturated and saturated zone model). These models need to be validated with observation data to confirm their reliability and accurate process representation \citep{Kennedy2023, Wagner2025}. 

The term model is used in two different ways in the presented literature: i) the gravity model, which represents the density variation over space (e.g. aquifer) and is used for the conversion of water storage change into gravity ($\Delta h$ into $\Delta g$) and ii) hydrological and hydrogeological model which calculates changes in water storage and fluxes through a defined area. Here, we use the gravity model (G-model) and hydrological model (H-model) to differentiate them. 

Gravimetry for measuring variation of water mass became of significant interest after the launch of the Gravity Recovery and Climate Experiment (GRACE) in 2002. GRACE can detect variations in surface water, soil moisture, groundwater, and snow, providing important data from regional to continental scales \citep{Ramillien2008}, in addition to monitoring sea level rise and mass changes of the solid Earth \citep{Chen2022}. GRACE revealed significant, large-scale depletion in aquifers in India, California and the Middle East, which are linked to human consumption, and gains due to climate-driven increases in sub-Saharan Africa and northern Amazon \citep{Rodell2023, ONeill2026}. Ground-based gravimetry from superconducting gravimeter stations can validate GRACE measurements \citep{Crossley2004,Hinderer2006,Kroner2009,Neumeyer2006, Neumeyer2008,Weise2009}

Beyond that, ground-based gravimetry is used to observe hydrological and hydrogeological processes. Gravity is influenced by regional (up to kilometers) and global hydrological mass variations \citep{Wziontek2009}, but regional influences are of interest for most applications. In these regional applications, the global influence needs to be removed from long-term gravity data. Due to its integrative character, multiple hydrological parameters can be captured with gravimetry over extended areas \citep{Hector2015}. \citet{Breili2009} showed that 58\% of the total gravity variation was attributed to snow cover. Highly sensitive superconducting gravimeters have been used to measure daily evapotranspiration fluxes of 1.7 mm \citep{VanCamp2016} or to evaluate radar-based precipitation estimates, as gravimeters capture rainfall at larger scales than rain gauges \citep{Delobbe2019}. Superconducting gravimeters can also capture infiltration processes \citep{Gettings2008}, whereas less sensitive gravimeters have been applied to monitor the larger mass changes of water in the saturated zone \citep{Hector2013, Pivetta2021, Pool1995}
Due to the current developments of new technologies in gravimetry and the resulting possibilities in hydrogeological applications, the objectives of this review are: 1) to describe different approaches for forward calculations of gravity changes caused by hydrological processes, 2) to provide an overview of gravimeters used for hydrological and hydrogeological measurements, 3) to report and summarize recent progress in instrument development and the potential it offers for field application, 4) to summarize applications in different hydrogeological settings and hydrological modelling and summarize their limitations, and 5) to identify directions for future research. 

\section{Principles of ground-based gravimetry}
According to Newton's law of universal gravitation, masses attract each other. The gravitational acceleration \textit{b} acting at a point is the vector sum of the attractive forces of all Earth's masses \citep[Chapter 3.1, page 66]{torge2025}. Due to the Earth's rotation, an additional centrifugal acceleration z acts on a mass at the Earth's surface \citep[Chapter 3.1.4, page 74]{torge2025}. The resultant of gravitational and centrifugal acceleration is termed gravity acceleration (gravity) g \citep[Chapter 3.1.4, page 76]{torge2025}:
\[
g = b + z
\]
Gravimeters measure gravity \textit{g}. Its magnitude varies globally (approx. 9.78 m/s$^2$ at the equator to 9.83 m/s$^2$ at the poles \citep[Chapter 3.2.1, page 78]{torge2025}), mainly due to the Earth's flattening and centrifugal force. The uneven mass distribution within the Earth and topography causes additional local variations. 

The gravity field can also be described as a scalar field, the gravity potential W, with $\mathbf{g} = \nabla W$ \citep[Chapter 3.1.4, page 76]{torge2025}. Surfaces of constant potential (W = const.) are called level surfaces or equipotential surfaces. Gravity $g$ is always perpendicular to these level surfaces \citep[Chapter 3.2.1, page 78]{torge2025}. The direction of \textit{g} defines the local plumb line (vertical), and the tangent plane defines the local horizontal. The geoid, representing the ocean's surface, if only impacted by  Earth's gravity and rotation alone, is a natural level surface.

Hydro-gravimetry primarily focuses on i) measuring temporal changes in gravity at a fixed point and ii) spatial differences between points, which is measured as the change in the vertical component of gravity acceleration. The SI base unit for gravitational acceleration is m/s$^2$. In this work, we use the following unit common in high-precision gravimetry \citep[Chapter 5.4.1, page 236]{torge2025}:
\begin{itemize}
  \item $1\ \text{nm/s}^2 = 10^{-9}\ \text{m/s}^2$
  \item $1\ \text{nm/s}^2 = 0.1\ \mu\text{Gal}$
\end{itemize}
Measurements are performed in the instrument's local coordinate system, defined by the local plumb line and the horizontal plane \citep[Chapter 2.5, page 53]{torge2025}. Two principal measurement techniques exist: 
1.	Absolute Gravimetry: Measures the total magnitude of gravity g directly by observing the length and time of a test mass in free fall \citep[Chapter 5.4.1, page 236]{torge2025} and 
2.	Relative Gravimetry: Measures differences in gravity ($\Delta g$) between different points or over time at one point. Typically, the displacement of a test mass suspended by a spring or the force required for compensation is measured \citep[Chapter 5.4.3, page 247]{torge2025}. Relative gravimeters require a reference (base station or previous measurement) and are subject to instrumental drift \citep[Chapter 5.4.3, page 248]{torge2025}.

\section{Data processing}
\subsection{Basics of data processing}
The redistribution of water masses causes the signal relevant for hydro-gravimetry. However, this signal is superimposed by other effects that also cause gravity variations and must be corrected for before the data can be used for hydrogeological interpretation and hydrological applications. Gravitational forces from the Moon and Sun cause periodic deformations of the solid Earth (solid Earth tides) and mass shifts in the oceans (ocean tides), leading to gravity changes of up to several 1000~nm/s$^2$ \citep[Chapter 3.8.2, page 120]{torge2025}. These are the largest non-hydrological effects. Changes in the distribution of air masses (atmospheric pressure variations) also cause gravity changes, typically in the range of tens of~nm/s$^2$ \citep[Chapter 5.4.1, page 238]{torge2025, Hinderer2006}. The shift of the Earth's rotation axis relative to the crust leads to small, long-period gravity changes ($<$1~nm/s$^2$)\citep[Chapter 3.8.1, page 119]{torge2025}. Local ground movements, seismic noise, and surface loading, that is, elastic deformation of the earth crust due to the weight of water \citep{Longuevergne2009}, can affect gravimetric measurements. An overview of models simulating the resulting signal of these processes is given in \citet{Mikolaj2019}.

After correcting the observed gravity variation for these effects, the resulting gravity residual $\Delta g_{\text{hydro}}$ ideally contains only hydrologically induced gravity changes and measurement noise \citep[Chapter 5.4.1, page 238]{torge2025}:

\[
g - \Delta g_{\text{earth}} - \Delta g_{\text{ocean}} - \Delta g_{\text{atmosphere}} - \Delta g_{\text{pole}} - \Delta g_{\text{drift}} \approx \Delta g_{\text{hydro}}
\]
with the individual corrections $\Delta g_{earth} $ being solid earth tides, $\Delta g_{ocean} $ being ocean tidal and non-tidal loading, $\Delta g_{atmosphere} $ being atmospheric pressure variations, $\Delta g_{pole} $ being polar motion, and $\Delta g_{drift} $ being the drift of the instrument itself. To obtain the local hydrological signal, corrections for large-scale hydrology have to be applied as well. 
\subsection{Forward calculation}
The forward gravity modeling in geodesy introduces equations to calculate the gravity effect of topography and the resulting anomaly from the Earth's gravity model \citep[Chapter 6.4.2, page 314]{torge2025, Gntner2017}. As topography influences water mass distributions on and below the Earth's surface, these approaches can be used to calculate the gravitational effect of water mass changes in the underground. 
In general, the gravitational force of a mass with distance r acting on a unit mass is described by
\[
g_z =
G \iiint \rho
\frac{z - z_m}{r^3}
\, dV
\tag{4.1}
\]
with G being the universal gravitational constant, $\rho$ being the density, $x_m, y_m, z_m$ being the coordinates of the sensor, $z-z_m$ being the vertical component of the distance vector $r= \sqrt{(x - x_m)^2 + (y - y_m)^2 + (z - z_m)^2}$, and dV the volume  \citep{Leirio2009,telford1990applied}.
A general formulation of hydrology-induced vertical attraction  $\Delta g$, including the spatially variable parameter $Sy$, is defined as follows:
\[
\Delta g = G \rho_w \int_{-\infty}^{+\infty} \int_{-\infty}^{+\infty} Sy(x, y) 
\left[ 
\sqrt{(x - x_m)^2 + (y - y_m)^2 + (h_f - z_m)^2} - 
\sqrt{(x - x_m)^2 + (y - y_m)^2 + (h_i - z_m)^2}
\right] \, dy \, dx
\tag{4.2}
\]
\begin{figure}[h]
\includegraphics[width=12cm]{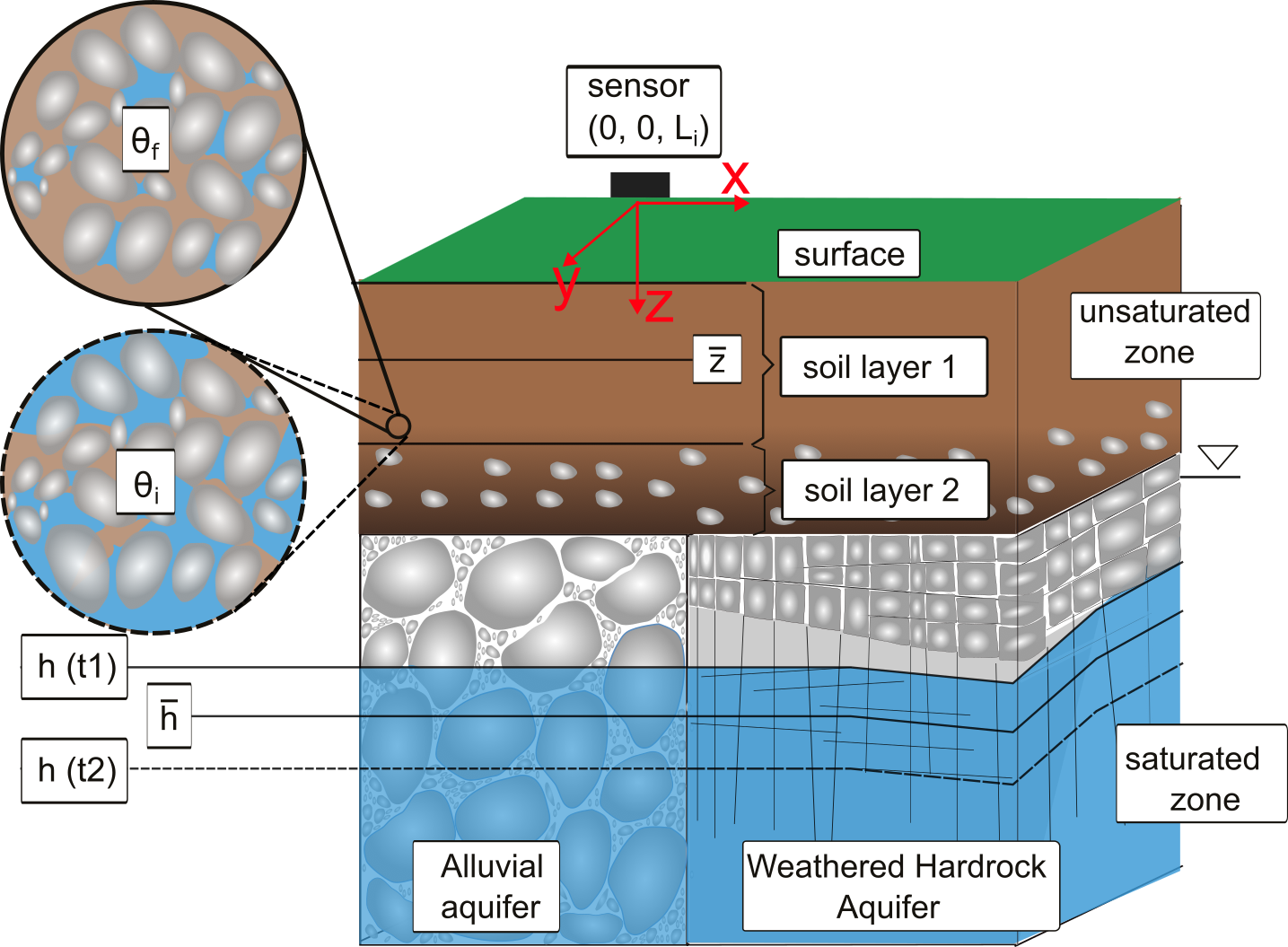}
\caption{In a hydrological model, the gravitational effect of changes in the water content in soil ($\theta$) is calculated in relation to the center of each soil layer, whereas the change of the groundwater table refers to the average groundwater table, after \protect\citep{Leirio2009}.}
\label{fig1}
\end{figure}
With $h_i$ and $h_f$ the initial and final water table, $Sy$ the spatially variable specific storage \citep{Leirio2009}. Based on this equation, \citet{Leirio2009} deduces equations for discrete spaces in a mesh-based H-model for both SZ and UZ. For a simple cubic cell with the same side length, the gravity change $\Delta g$ in the vertical direction caused by groundwater storage change (GWSC) can be calculated with the closed-form of the point-mass approximation for a cell whose mass is assumed to be concentrated at its center, thus the real dimensions of the volume is ignored: 
\[
\Delta g _{hydro}
= G\,\rho_{\mathrm{w}}\,Sy\,\Delta h \,
\frac{L_i-\bar h}
{\left[(L_i - \bar h)^2 + x^2 + y^2\right]^{3/2}}\Delta x \, \Delta y
\tag{4.3}
\]

with $\rho_w$ being the density of water, $Sy$ the specific yield, $dh$ the different water level between two time steps,  $L_i$ the z coordinate of the instrument, $\bar h$ the mean water level (figure \ref{fig1}), $\Delta x$, $\Delta y$ the lengths of the cell sides, and x, y, z being the vector components of the  point-mass relative to the instrument’s position 
$d = \sqrt{x^{2} + y^{2} + z^{2}}$.  For the UZ, the point-mass approximation is applied to calculate $\Delta g$ caused by a difference in soil moisture (figure \ref{fig1}):
\[
\Delta g_{hydro}
= G\,\rho_{\mathrm{w}}\,(\theta_f-\theta_i)\,\Delta z\;
\frac{L_i-\bar z}{\bigl(x^2+y^2+(L_i-\bar z)^2\bigr)^{3/2}}\Delta x \Delta y
\tag{4.4}
\]
with $\theta_i$ and $\theta_f$ being the initial and final soil moisture, respectively, and $\bar z$ the depth of the soil layer center. The total $\Delta g$ is the sum of $\Delta g$ from the individual soil layers \citep{Leirio2009}. Other forward calculations are applied in hydrological studies, as integrals of the mass-point formula over varying domains. 

\begin{figure}[t]
\includegraphics[width=12cm]{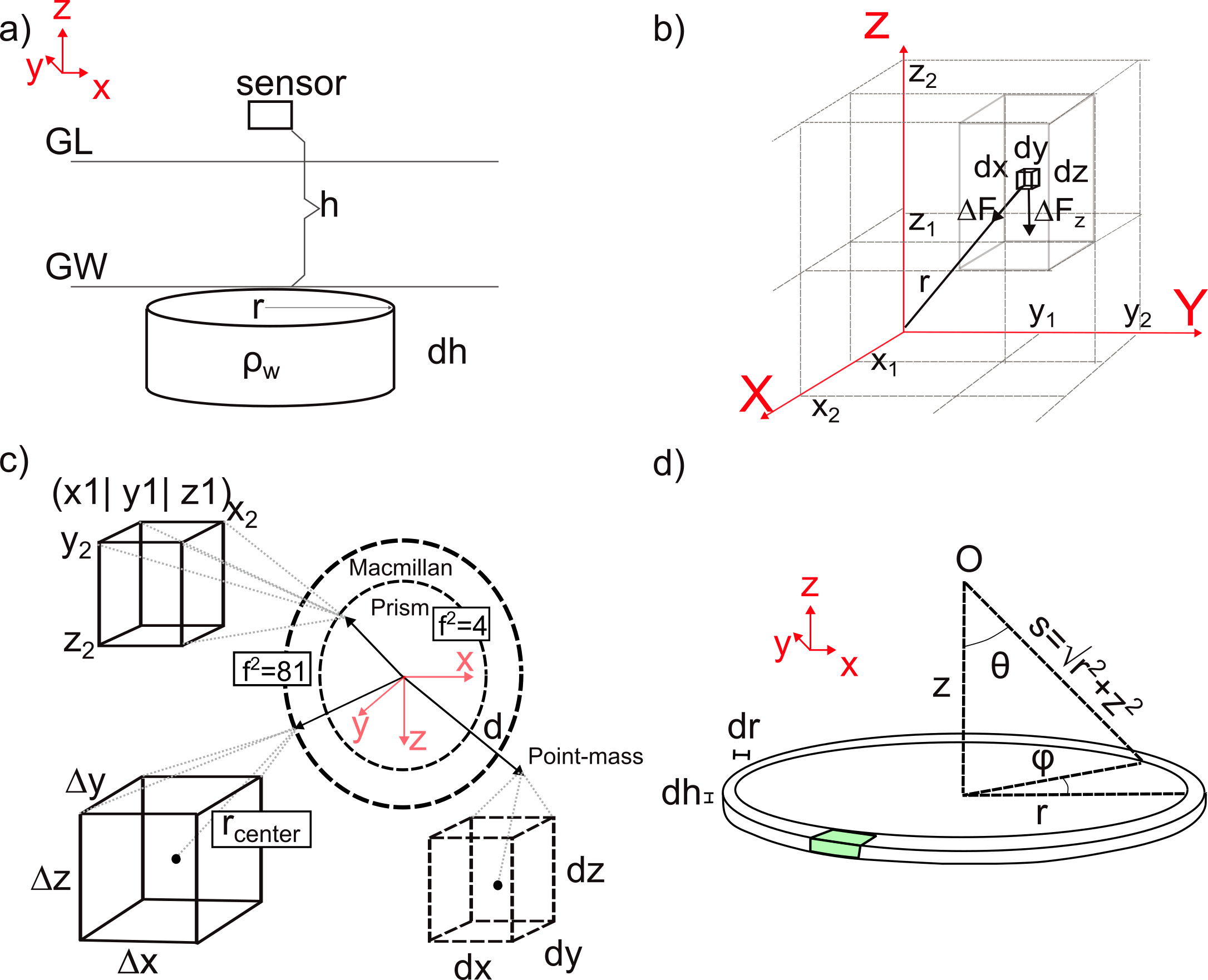}
\caption{Different solutions to the gravitational potential of the topography: a) cylinder approach with sensor above ground level (GL) \protect\citep{Damiata2006}, b) prism approach for calculating the vertical gravity component caused by a rectangular prism \protect\citep{Nagy1966}, conceptual representation based on \protect\citep{Nagy1966}, c) prism -, point mass -, and MacMillan approach for different distances to the sensor \protect\citep{Leirio2009}, conceptual representation based on \protect\citep{Leirio2009}, d) $\Delta g$ at point O due to GWSC as a thin layer $\Delta h$ in GRAVi4G, from \protect\citep{Halloran2022}, published under CC BY-NC-ND 4.0}
\label{fig2}
\end{figure}

For a limited lateral extent of the GWSC and in systems that are radially symmetric \citep{Damiata2006}, the conversion of GWSC into a homogeneous, finite cylinder can be calculated as follows:
\[
\Delta g = 2\pi \rho_w  G Sy \left[ d + \sqrt{r^2 + h^2} - \sqrt{r^2 + (h + dh)^2} \right]
\tag{4.5}
\]
with $r$ and $dh$ being the cylinder's radius and height, respectively, and h the distance of the source (sensor) to the cylinder along the axis through the cylinder’s center figure (\ref{fig2} a) \citep{HeiskanenMoritz1967, Gntner2017}. This approach is suitable for the analysis of a pumping test, due to the radially symmetrical drawn-down cone \citep{gonzalezquiros2014}. When the lateral extent of the cylinder is unlimited, the GWSC is described by the Bouguer plate \citep{torge2025, Gntner2017}: 
\[
\Delta g = 2\pi \rho_w G Sy \Delta h
\tag{4.6}
\]
In this approximation, $\Delta g$ is only dependent on $\Delta h$, $\rho_w$, and $Sy$, 1~m GWSC results in a $\Delta g$ of 419~nm/s$^2$ \citep{Arnoux2020, Christiansen2011c, Gntner2017, Leirio2009}. This approach can be applied when the lateral extent of the source of gravitational attraction (groundwater body in our case) is much larger than the distance between the measuring point and the center of the groundwater body. Hence, it is more appropriate for shallow, unconfined aquifers with a broad, homogeneous water level change \citep{Halloran2022, Pool1995}. \citet{Chen2020} describes the Bouguer approximation as the only plausible method for calculating $\Delta g$, when the extent of an aquifer is unknown. It is often used in studies investigating porous aquifers, due to their homogeneous nature \citep{Chen2020, Christiansen2011b, El-Diasty2016, Pendiuk2020, Pool2008, Pool1995}. 

The spatial variability of the correlation between $\Delta g$ and $\Delta h$ is considered in the prism \citep{Nagy1966} and the polyhedron approach \citep{Barnett1976}. \citet{Nagy1966, Nagy2000} developed a closed expression for the vertical gravity component caused by a prism acting on any point outside or at the boundary of the prism. As it is possible to form any kind of body with prisms of varying size and density, this prism approach or Forsberg equation \citep{Forsberg1984} can be applied to calculate the vertical component of gravity caused by varying elements (figure \ref{fig2} b). 

\[
F_z = G \rho \left[ 
x\ln(y + d) + y \ln(x + d) - z \arcsin \left( \frac{z^2+y^2+yd}{(y+d)\sqrt{y^2 + z^2}} \right)
\right] \Bigg|_{x_1}^{x_2} \Bigg|_{y_1}^{y_2} \Bigg|_{z_1}^{z_2}
\tag{4.7}
\]
With $F_z$ being the vertical gravity component and d the distance between prism and measuring point (zero) in a Cartesian coordinate system and $x_1, x_2, y_1, y_2, z_1, z_2$ defining the boundaries of the prism. For a hydrological model $\rho = \rho_w \cdot Sy$ applies \citep{Leirio2009}. 

In summary, \citet{Leirio2009} presents three approaches to specifically calculate the temporal gravity change ($\Delta g$) resulting from the mass changes of a finite-difference H-model, which is particularly suitable for post-processing of the output of finite-difference hydrogeological models. The three different approaches are applied according to the distance between the sensor and the attracting mass and the cells dimension (figure \ref{fig2} c).  The normalized distance $f^2 = \frac{r^2}{dr^2}$ is defined to select the formula for each space. With r being the distance from the cell center to the sensor, while dr is the cell size. If $f^2$ < 4, the prism formula is applied; if 4 < $f^2$ < 81, the MacMillan (Appendix \ref{sec:mcm}) formula; and if $f^2$ > 81, the point-mass formula is applied. 
 \citet{Barnett1976} describes an analytical integral over triangular facets to compute gravitational fields of an arbitrary 3-D body represented as a polyhedron made of triangles, which could be applied for finite-element meshes.

For mountainous regions with significant relief, the Bouguer plate approximation is not valid. A recently developed software improves the conversion of $\Delta g$ to GWSC for non-flat terrain \citep{Halloran2022}: Gravi4GW is a Python package to estimate the GWSC from the change in gravity, while considering topography. Storage changes occurring at the land surface are described as a thin layer distributed across the local topography. This approach calculates $\Delta g$ for an infinitesimal mass element of a horizontal layer with the thickness $\Delta h$. The tool provides a conversion factor $\beta$ (eq. 4.8), which is the absolute value of the vector $\vec{\beta}$ that includes both vertical and radial components, thereby accounting for the difference in conversion factors for various terrain slopes. This is because $\Delta g$, caused by GWSC, does not necessarily occur in the vertical direction (plumb line). $\vec{\beta}$ is defined as the change in gravity caused by the water column change $\partial h$: 

\[
\vec{\beta} = \frac{\partial \vec{g}}{\partial h}\tag{4.8}
\]
With the assumptions of i) a planar water table, ii) a thin $\Delta h$, iii) uniform density, and iv) GWSC only impacts the vertical component ($\beta_z$ = $\beta$), the value of $\beta$ is calculated as follows:
\[
\beta = 2\pi \rho_w  G Sy\int_0^{r_0} \frac{r z}{(r^2 + z^2)^{3/2}} \, dr = 2\pi G \rho \left[1 - \frac{1}{\sqrt{1 + \left(\frac{r_0}{z}\right)^2}} \right]
\tag{4.9}
\]
With $r_0$ being the radial distance of the area and z the vertical distance between the sensor and the attracting point figure \ref{fig2} d). The error occurring from the deviation of the theoretically infinite extension of the gravitational effect and the integration over a finite radius $r_0$ is estimated as follows:
\[
\varepsilon = \frac{1}{\sqrt{1 + \left( \frac{r_0}{z} \right)^2}} \approx \frac{z}{r_0}
\tag{4.10}
\]
The error decreases linearly with increasing radius \citep{Halloran2022} and supports the results from \citet{Gntner2017} and \citet{Leirio2009} that over 90\% of the gravity signal is captured, when $r_0$=10 z. Following the assumptions i) and ii), the tool is applicable for low water storage changes, and a water table following the terrain slope. 

 \subsection{Gravity inversion}
Following the forward calculation, the density distribution of the subsurface is determined via gravity inversion, which aims to find a model that fits the observed data. The data inversion process of geophysical data is summarized in \citet{binley2015}. 
The inversion of gravity data can be applied to either determine the geometry of a subsurface interface or the 3D density variation \citep{Zhong2022ConstrainedGravity}.  In this second approach, the subsurface is divided into voxels, and their density is the unknown, targeted in the inversion \citep{kaban2016}. How well the density model explains the measured gravity data is described by an objective function, which is minimized during inversion \citep{kaban2015cratonic, kaban2016}. 
 \[
\min \left( \left\| A\rho - g_{\mathrm{res}} \right\|^2 + \alpha \, \Omega(\rho) \right)
\]
with A being the operators converting densities into gravity, $g_{res}$ the residual gravity, and $\Omega$ the regularization term, of which $\alpha$ determines the strength of the regularization. The regularization term is a penalty term that introduces prior knowledge to the least square function \citep{tikhonov1977illposed, KARL2005183, Zhong2022ConstrainedGravity}. \citet{Zhong2022ConstrainedGravity} adds a cross-gradient term to the objective function in addition to the regularization term. The regularization term includes an a priori (reference) model, whereas the cross-gradient term incorporates information from other geophysical measurements by promoting consistency between different physical property models.   Various approaches to integrate multiple geophysical observations into the objective function exist and are reviewed by \citet{colombo2018}. 

As water distribution is related to density distributions in the subsurface, these approaches can be applied to constrain H-models. The benefit of including hydrological models in the inverse geophysical approach has already been described by \citet{Hinnell2010}. The coupled or joint approach reduced the uncertainty of predictions and hydrological parameters estimations significantly, compared to the independent geophysical inversion \citep{Hinnell2010}. Also, additional datasets can be 
included in the coupled inversion to achieve a multiobjective optimization \citep{binley2015}. 
 
\section{Devices}
The main types of gravimeters used in hydrogeological studies are absolute gravimeters (AG), relative gravimeters (RG), and superconducting gravimeters (SG), a comparison of their characteristics is given in table \ref{table0}. 

Additionally, quantum and MEMS gravimeters are mentioned as a promising future technology. Key instrument characteristics that determine the detectability of mass changes are precision and accuracy (see Glossary). 
The reported accuracy depends on the environmental noise at the specific site. 
\subsection{Absolute gravimeters}
Modern AG uses the free-fall principle of a test mass in a vacuum \citep[section~5.4.1]{torge2025}. The position of the test mass, which is often a corner-cube reflector \citep{Faller1965}, is tracked with high precision using laser interferometry during its fall \citep{Niebauer1995}, while the corresponding time points are registered with an atomic clock. From the distance-time relationship z(t), considering the gravity gradient \citep{Niebauer1995}, the gravity value g is determined by least-squares adjustment of numerous points measured during a single drop \citep{Niebauer1995}. 
Raw measurements must be corrected for various effects, notably tides (Earth \& ocean), atmospheric pressure variations, and polar motion \citep[section~5.4.1]{torge2025}. An existing vertical gravity gradient must also be accounted for \citep{Niebauer1995}. The achievable accuracy strongly depends on site conditions like stability and microseismicity \citep{VanCamp2017} but typically reaches 10--50~nm/s$^2$ after averaging over hundreds to thousands drop experiments over 1--3 days \citep{Crossley1999}. Systematic differences, so-called "offsets" between different instruments, can also be of this magnitude \citep{Francis2001}.

\subsubsection{Quantum gravimeters (atom interferometers)}
A more recent development involves absolute quantum gravimeters (AQG) based on atom interferometry. Clouds of laser-cooled atoms serve as test masses in free fall \citep[section~5.4.2]{torge2025}. Their acceleration is measured interferometrically by interaction with precisely timed laser pulses, acting as beam splitters and mirrors for the atom waves \citep{Kasevich1991, Peters2001}. The resulting phase shift in the interferometer is directly proportional to g \citep{Peters2001}. AQGs potentially combine the advantages of absolute gravimeters (drift-free, SI traceability) with high sensitivity and rapid measurement rates \citep{Freier2016GAIN}. They have no moving mechanical parts \citep{Peters2001}. Current transportable devices achieve accuracies in the $\mu$Gal ($nm/s^{2}$) and stabilities in the sub-$\mu$Gal range \citep{Freier2016GAIN, Menoret2018}.

\subsection{Relative gravimeters}
Most commercial field geophysical surveys are executed with traditional RG that rely on mechanical or optical principles. These include spring-based gravimeters, SGs, and micro-electromechanical systems (MEMS) gravimeters. Spring based RG are among the most widely used instruments due to their robustness and cost-efficiency. In a spring-based gravimeter, a test mass is suspended by a high-precision spring within the gravity field. Changes in gravity lead to a small change in the spring's length or the test mass's position \citep[section~5.4.3]{torge2025}. This displacement is either measured directly or, more commonly, compensated by a counteracting force, i.e. mechanically via a measuring screw or electronically via feedback \citep{Harrison1984}. The displacement or the compensation force is approximately proportional to the gravity change \citet[Chapter 5.4.3, page 247]{torge2025}. Therefore, an RG does not measure directly in gravity units, and the readings (e.g., scale units or Volts) must be converted to nm/s$^2$ using a calibration function that is often approximated by a linear scale factor, potentially with non-linear/periodic terms. 
To achieve high sensitivity, many modern instruments utilize astatization or the "zero-length" spring concept \citep{LaCoste1934}, as described in \citet[Chapter 5.4.3, page 248]{torge2025}. However, they are prone to significant instrumental drift and require regular calibration against a known reference point, typically using an absolute gravimeter, to maintain measurement accuracy \citep{Shettell2024}. \citet{LewisRodgers2018} performed field measurements with Scintrex CG5 and CG6 spring-mass gravimeters and detected a drift up to $0.7 \times 10^{-6} m/s^2$ over 3 hours, which needed to be calibrated. 
\subsubsection{Superconducting gravimeters}
SG represents the most precise and stable class of relative gravimeters, but they are typically stationary. Therefore, they are particularly suitable for continuous long-term gravity observations \citep{Fang2024}. They are, however, largely limited to fixed installations due to their size and infrastructure needs. A superconducting sphere levitates stably in the magnetic field of superconducting coils, maintained at cryogenic temperatures with liquid Helium \citep[section~4.2]{torge2025}. Gravity changes cause vertical displacements of the sphere, which are detected capacitively and compensated for by a feedback voltage. This voltage is directly proportional to $\Delta g$ \citep{Prothero1968, Sasagawa1989}. SG offers the highest sensitivity and resolution (<0.01 nm/s$^2$  or <0.001 $\mu$Gal \citep{Warburton2010}) and exceptional long-term stability with extremely low drift of ~1 µGal/year \citep{Hinderer2007}. 

\begin{table*}[htbp]
\caption{Overview of gravimeters applied in hydrology: expected stability, accuracy, and reported application. AG: absolute gravimeter; RG: relative (spring) gravimeter; SG: superconducting gravimeter; AQG: absolute quantum gravimeter; MEMS: micro-electro-mechanical system.}
\scriptsize
\setlength{\tabcolsep}{4pt}
\renewcommand{\arraystretch}{1.3}
\begin{tabularx}{\textwidth}{@{}p{1.8cm} *{5}{>{\raggedright\arraybackslash}X}@{}}
\tophline
 & \textbf{AG (cube)} & \textbf{RG (spring)} & \textbf{SG} & \textbf{AQG (atom)} & \textbf{MEMS} \\
\middlehline
Short-term stability
 & $100$ nm\,s$^{-2}$/$\sqrt{\text{Hz}}$ up to ${\sim}300$\,s (a)
 & $600$ nm\,s$^{-2}$/$\sqrt{\text{Hz}}$ up to ${\sim}300$\,s
 & $1$ nm\,s$^{-2}$/$\sqrt{\text{Hz}}$
 & $100$ nm\,s$^{-2}$/$\sqrt{\text{Hz}}$ up to ${\sim}10^{5}$\,s
 & $34$ nm\,s$^{-2}$/$\sqrt{\text{Hz}}$ up to ${\sim}30$\,s (c) \\
Long-term stability
 & $10$ nm\,s$^{-2}$ (d)
 & drift ${\sim}200$ nm\,s$^{-2}$/d
 & ${\sim}0.01$ nm\,s$^{-2}$/d (e)
 & $0.5$ nm\,s$^{-2}$ (f)
 & drift $860$ nm\,s$^{-2}$/d (c) \\
Accuracy
 & $10$--$50$ nm\,s$^{-2}$ (e)
 & n/a (relative)
 & n/a (relative)
 & $30$--$40$ nm\,s$^{-2}$ (a)
 & n/a (relative) \\
Specifics
& SI traceability
 & compact, portable
 & SI traceability, continuous high-frequency recording; special conditions
 & SI traceability, continuous high-frequency recording; specialconditions
& cost-efficient \\
Application
 & reference stations; calibration; long-term
 & network surveys; monitoring; cost-effective; compact; local gradient
 & stationary; local studies; time-lapse measurements
 & transportable; network and time-lapse surveys
 & long-term time-lapse gravimetry; monitoring networks \\
\bottomhline
\end{tabularx}
\belowtable{
(a)~\citet{gillot2016lne_syrte_gravimeter};
(b)~\citet{VanCamp2017};
(c)~\citet{Gao2026ChipScaleGravimeter};
(d)~\citet{Niebauer1995};
(e)~\citet{torge2025};
(f)~\citet{Freier2016GAIN}.
}
\label{table0}
\end{table*}

\subsubsection{Micro-Electromechanical systems}
In contrast, relative MEMS (micro-electromechanical systems) gravimeters offer a more cost-efficient and smaller solution, while providing comparable sensitivity and stability. MEMS gravimeters can also measure changes in gravity by monitoring the displacement of a test mass on a spring. The system measures the movement of the test mass via overlapping comb capacitors, whose capacitance changes with each displacement of the mass. This change in capacitance modulates the output current, which is converted into a position signal via a transimpedance amplifier topology \citep{prasad2022}. The advantage of combining multiple sensors into an array of RG, while connecting them to an AQG has been shown for volcanic monitoring, as the spatial resolution is increased and continuous measurements are facilitated \citep{Carbone2020_NEWTONg}. State-of-the-art MEMS reach 0.1 µGal after 1s measurement time and then start to drift. They demonstrated a direct detection of Earth tides after a removal of the high-order drift. The advantage of these devices is that they can be produced using semiconductor processes \citep{Wu2019}.

\subsection{Advances in ground-based quantum gravimeters for field application}
\begin{figure}[t]
\includegraphics[width=12cm]{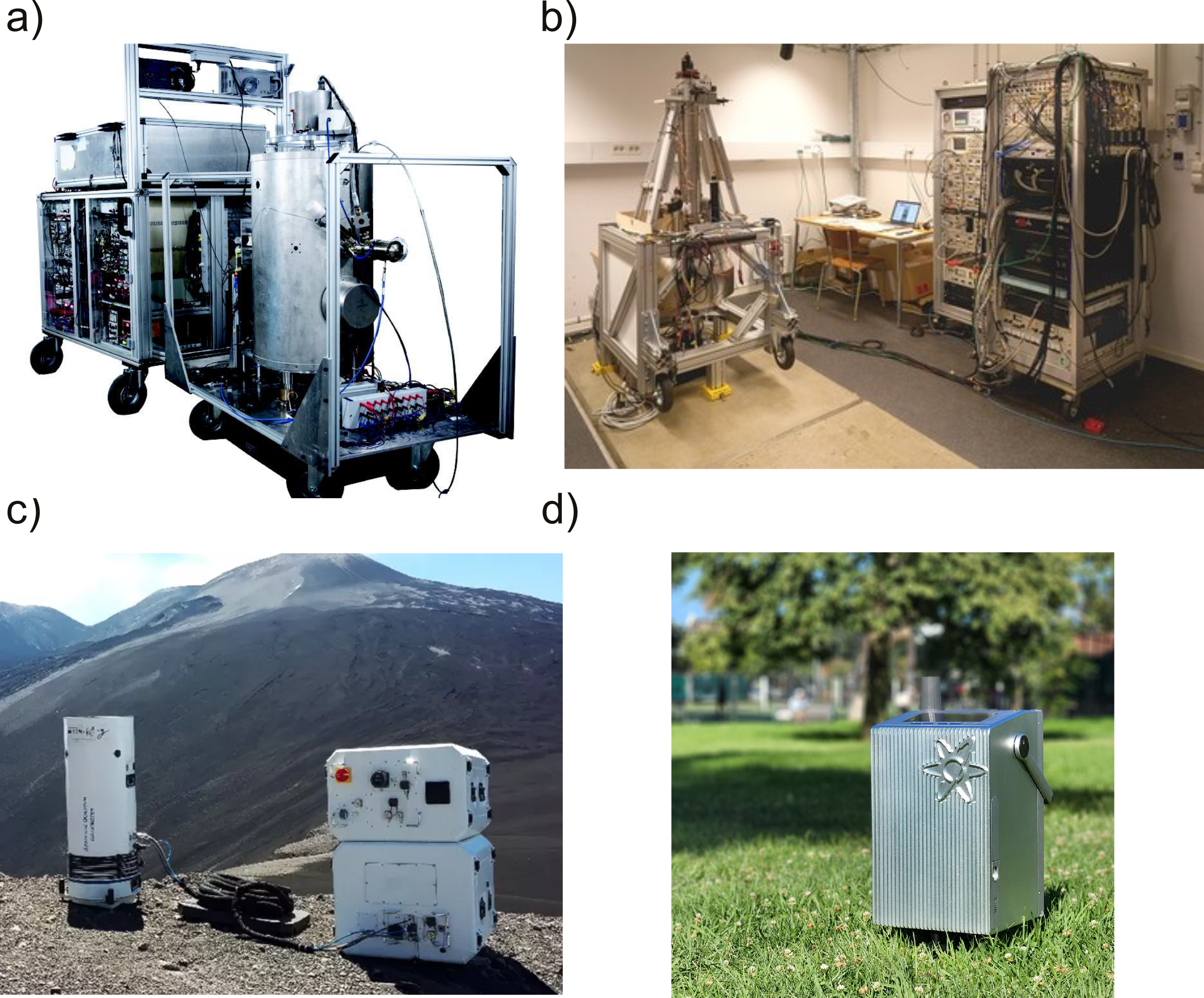}
\caption{Typical transportable static atomic quantum gravimeters include state-of-the-art gravimeters for research a) LNE-SYRTE in France \protect\citep{Farah2014} (photo used with permission by Sébastien Merlet); b) Humboldt University in Germany, and for field investigations c) Exail in France \protect\citep{ExailQuantumGravimeters} (photo used with permission by Exail); d) Nomad Atomics in Germany and Australia. An extended overview of transportable static atomic gravimeters is given in \protect\citet{Fang2024}.}
\label{fig3}
\end{figure}

While most commercial gravity surveys continue to rely on traditional RGs, recent advances have enabled the deployment of AGs and, more recently, quantum-based gravimeters in field environments. These instruments (figure \ref{fig3} c), d)), previously confined to laboratory or observatory settings, now demonstrate sufficient robustness and portability for use in varying conditions. In 2015, the first commercial gravimeter based on laser-cooled atoms, the Absolute Quantum Gravimeter (AQG), was launched \citep{Arnal2023_QuantumGravitySensors}. 

Additional field measurements have been conducted during an evaluation by \citet{Cooke2021}, in which the AQG B01 was tested over several weeks under diverse environmental conditions. In a low-noise environment, the instrument exhibited repeatability better than 1 $\mu$Gal after 1 h. As part of the NEWTON-g project on Mount Etna, a hybrid approach that integrates MEMS-based relative sensors with an AQG was deployed. The AQG system demonstrated continuous operational robustness in the challenging volcanic environment, successfully detecting gravity variations linked to volcanic activity in the range of ~1 to ~100 $\mu$Gal, sensitive enough to monitor subsurface mass changes such as magma movement \citep{Carbone2020_NEWTONg}. 
The emergence of commercially available quantum gravimeters marks a significant turning point in the evolution of field gravimetry. Historically, high-precision absolute measurements were often restricted to laboratory-based settings or required specialized field-portable instruments (e.g., the A10), whereas the development of AQGs introduces more robust, autonomous, and field-deployable quantum gravimeters.

This offers several advantages: i) drift-free measurements with no need for repeated calibration or reference station, ii) high sensitivity with long-term stability below $\sim$1~$\mu$Gal over days to weeks, and iii) operability outdoors without a complex setup. While practical applications require compact, portable gravimeters, most traditional AGs are bulky, weighing around 30~kg, and require several supporting components such as a separate laser system, control electronics, and power supplies \citep{Menoret2018}. As described by \citet{Veryaskin2022}, several companies currently develop commercially available AQG, with a focus on compact, low-cost cold-atom gravimeters, designed to be moved and operated by a single person in the field \citep{Cordier2026,Cooke2021,  Wu2019, Veryaskin2022}. Nomad Atomics recently demonstrated the operation of their device in an underground mine \citep{Cordier2026}, whereas Exail's AQG was deployed near volcanic craters (\ref{fig3} c) to detect mass changes of active volcano \citep{Micollier2020}.

\section{Applications in Hydrology}
Gravimetry has been applied in varying hydrological or hydrogeological studies and table \ref{table1} summarizes hydrological processes and the resulting change in gravity. A study investigating the impact of hydrogeological parameter uncertainty on gravity change during pumping tests describes the difference between homogeneous and heterogeneous aquifers. In homogeneous aquifers, gravity changes are smoother, more predictable, and primarily influenced by a few parameters, especially hydraulic conductivity (Ks) and Sy. In heterogeneous aquifers, gravity responses are more variable, uncertain, and sensitive to spatial structure, thereby carrying the information of all uncertain parameters \citep{Maina2021}. Use cases of hydro-gravimetry in varying geological settings are summarized in porous (homogeneous) and weathered hard rock (heterogeneous) settings. 
\begin{figure}[t]
\includegraphics[width=12cm]{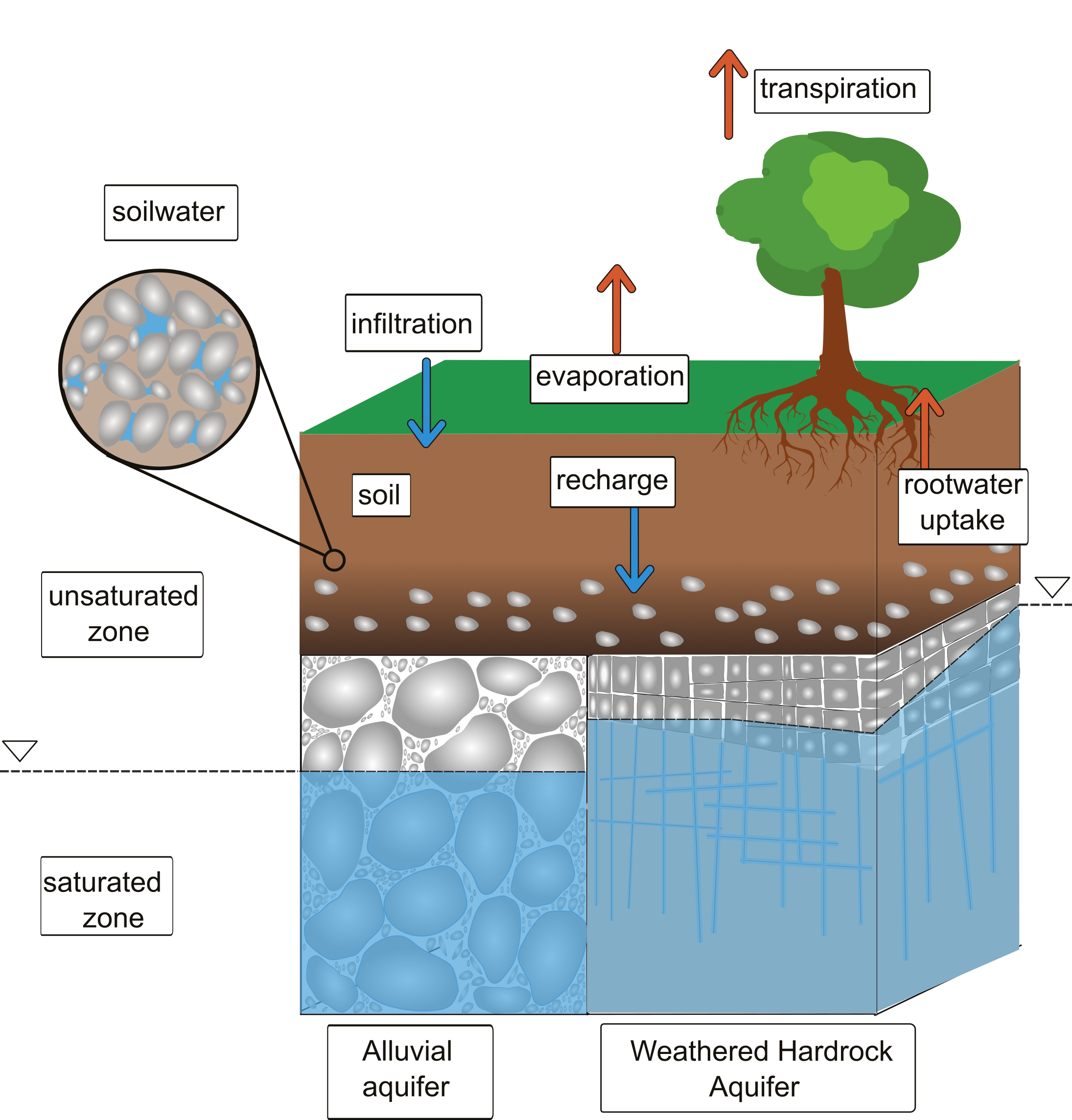}
\caption{Conceptual image of porous and weathered hard rock, including the unsaturated zone and water fluxes.}
\label{fig4}
\end{figure}

\subsection{Application of hydro-gravimetry in alluvial aquifers}
Studies applying hydro-gravimetry in a porous aquifer often aim at estimating $Sy$ \citep{Blainey2007, Chen2020, Christiansen2011c, El-Diasty2016, Hsiao2021, Pendiuk2020, Pool2008, Pool1995, gehman2009, herckenrath2012, seraphin2018}, and groundwater storage changes (GWSC) \citep{El-Diasty2016, Hernndez-Snchez2021}. The approach to estimate $Sy$ varies across the studies \citep{Hsiao2021, Pool2008, Pool1995} and \citep{Chen2020} calculate $Sy$ as a ratio of $\Delta g$ to the water level variations ($\Delta h$) with the following equation, which is based on the Bouguer approximation: 
\[
Sy = \frac{\Delta g}{2\pi \rho_{\mathrm{w}} G (Z_1 - Z_2)}
\tag{5.1}
\]
with G = 6.67E-11 m$^3$/kgs$^2$ and $\rho_w$ = 1000 kg/m$^3$, $Z_1$ and $Z_2$ being the water levels of two surveys and $\Delta g$ the corresponding difference in gravity. \citep{Pool1995} investigated a groundwater level change ($\Delta h$) of up to 17.7 m, which caused a $\Delta g$ of 1578~nm/s$^2$ within a network measurement, a closed loop of two base stations and six locations near wells with a relative RG. Even though the instrumental drift and errors caused by changes between the reference station and field station result in significant noise, the gravity survey during different seasons and years helped to estimate $Sy$ and GWSC \citep{Pool1995}. In a second study, repeated, paired measurements of water level and $\Delta g$ were fitted by linear regression to calculate $Sy$ \citep{Pool2008}. Results in this study indicated different conditions of the underground. Unrealistic $Sy$, larger than 0.35, indicated the contributions of water in the UZ to $\Delta g$. Also, confined conditions could be identified at locations with larger water level changes (> 4m) and low $\Delta g$ (< 10 $\mu$Gal) \citep{Pool2008}. 
\citep{Chen2020} measured the difference in groundwater table during the wet and dry seasons at 10 different sites. The depth to groundwater is between 10-20 m below surface level (b.s.l.), and the difference in groundwater level (peak to trough) was between $\Delta h$ = 0.5-8 m. Changes in soil moisture were only calculated for three sites, and the resulting gravity signal is lower than the error of the applied gravimeter (FG5). Changes in gravity were measured in sessions of 12-16 h, which comprises sets of 30min with an absolute gravimeter. In each set, 100 drops are conducted. The gravity signal was averaged over the whole session. This increased the error of gravimetric measurement, as groundwater level changes up to 0.2 m per day and thus influences the average gravity value by about 1.7 $\mu$Gal. Chen et al. (2020) described a simple method for the estimation of $Sy$ efficiently while considering errors caused by the Bouguer approximation, error of the gravimeter, and precision of the piezometer (sensors measuring pore water pressure).

A different approach was published by \citet{Pendiuk2020} in the same year. The hydrological and hydrogeological contributions to gravity variations were modelled, and the resulting time series is used to inversely estimate $Sy$ by finding the minimum of the objective function between modelled and measured gravity residuals. The approach consists of the integration of formulas representing the gravity signal from each hydrogeological compartment. It includes equations to calculate the gravitational response caused by i) groundwater table decline, ii) evapotranspiration, and rainfall infiltration in the UZ. The gravity signal caused by soil and porous media is approximated by the infinite Bouguer approximation and varying density of water and soil matrix. The influence of the groundwater table fluctuation on the gravity signal is then described as the difference of the gravity signal of two different points in time, as presented in eq. 5.1. The influence of the UZ above the groundwater table is described by a function describing i) time variable rainfall and ii) dynamic processes of water storage in the soil and water losses as evapotranspiration. The addition of both compartments results in the following equation:	
\[
\Delta g = 2\pi G \rho_{\mathrm{w}}
\left[
Sy (z_1 - z_2) + r(t_r)\bigl(f(t_2) - f(t_1)\bigr)
\right]
\tag{5.2}
\]
with $r(t_r )$ representing the function of rainfall, while $f(t_2 )$ is the function describing infiltration and evapotranspiration at times $t_1$ and $t_2$\citep{Pendiuk2020}.
In the work of \citet{El-Diasty2016}, GWSC is detected by combining hydro-gravimetry and soil moisture change (SMC), as gravimetry only gives information about the total water storage change (TWSC). $\Delta g$ was inverted into TWSC according to the Bouguer approximation. SMC data were taken from a land surface model (LSM) covering the whole study area. The LSM soil moisture data was validated by comparing it with the $Sy$ determined during geological studies. The annual water level change $\Delta h$ from summer to winter is 0.6m, and the UZ is 11.25m deep; nonetheless, the GWSC could be estimated for an area of 112 km$^2$. The measurements were conducted with an AG and a network measurement of RG \citep{El-Diasty2016}. Even though errors introduced by the LSM, which is only an estimation of actual soil moisture, were not considered, and the estimated GWSC could not be directly validated, the study seems promising to apply hydro-gravimetry for GWSC estimation. 
\newline
Gravimetry to quantify GWSC was applied in an alpine catchment, where meltwater flows through different geological features (e.g., talus and moraines) \citep{Arnoux2020} with varying Ks \citep{Caballero2002}. Alpine regions are characterized by steep topography and are located above the tree line \citep{Arnoux2020}, thus difficult to monitor. Two gravity surveys were conducted to estimate GWSC in different geological features. The combination of gravimetry with isotope and thermal tracers could help to understand the flow regime in the catchment, even though GWSC was overestimated by gravimetry. This might be caused by the simplified conversion of $\Delta g$ to $\Delta h$ using the Bouguer approximation.
\newpage
\begin{sidewaystable*}[htbp]
\caption{Examples of gravity changes ($\Delta g$) associated with hydrological processes.}
\resizebox{\textwidth}{!}{%
\begin{tabular}{l l c c p{2.5cm} p{6cm} l}
\tophline
Parameter & Unit of $\Delta h$ & $\Delta h$ & $\Delta g$ [nm s$^{-2}$] & Instrument & Specifics & Source \\
\middlehline

Drawdown pumping test & m & 10 & 270 &  & Synthetic pumping test with pumping rate of 0.06309 m$^3$/s & \citep{Leirio2009} \\

Evapotranspiration & mm/day & 1.7 & $1.0\times10^{-8}$ & SG C021 & Oak–beech forest, evapotranspiration over an area of 50 ha & \citep{VanCamp2016} \\

WSC of soil & \% & 13 & 27 & SG C021 & Soil moisture measured with soil moisture probes & \citep{VanCamp2016} \\

Infiltration MAR pond & m/day & 0.35 & 750 & SG iGrav-007 & Infiltration investigated over a period of 1 month, SG positioned at the pond & \citep{Kennedy2016} \\

Rainfall & mm/h & 75 & 6 & SG C021 & SG installed 48 m below surface & \citep{Delobbe2019} \\

Rainfall & mm & 1 & 1.96 & SG iGrav-007 & Derived admittance coefficient, signal referenced to rainfall stored in the soil above the gravimeter & \citep{Luan2023} \\

Rainfall & mm/h & 6 & $\sim 2$ & SG CD-034 & Immediate signal referenced to rainfall stored in the soil above the gravimeter & \citep{Weise2018} \\

Water table in reservoir & m / 1.15 h & 0.68 & 270 & Scintrex CG-5 & Experiment: outlet of 20 m $\times$ 30 m indoor basin causing $\Delta h$ & \citep{Christiansen2011b} \\

Water table in reservoir & mm/day & 1 & 2.2--2.6 & SG C021 & Distance to gravimeter 3–6 km & \citep{VanCamp2006} \\

WSC groundwater & m & 3 & 170 & SG-060, FG5 & Crystalline aquifer & \citep{Hector2015} \\

WSC groundwater & m & 1 & 6.5 & SG iGrav-007 & Derived admittance coefficient, weathered hard rock aquifer & \citep{Luan2023} \\

WSC groundwater & m & 3 & 100 & AG FG5, SG 030 iGrav006 & Weathered hard rock aquifer & \citep{Hector2013} \\

WSC groundwater & m & 1 & 27 &  & Admittance coefficient, weathered hard rock aquifer & \citep{Hokkanen2006} \\

WSC groundwater & m & 0.6 & 50 & SG 038 & Porous aquifer & \citep{Pendiuk2020} \\

\bottomhline
\end{tabular}
}
\belowtable{}
\label{table1}
\end{sidewaystable*}
\newpage
\subsection{Application of hydro-gravimetry in crystalline aquifers}
The level of heterogeneity of the geological setting must be considered when applying gravimetry for estimating GWSC. A weathered hard rock aquifer can be homogeneous in the upper, weathered layer, but tends to be anisotropic in the deeper, fractured part \citep{Murty2002}. Due to the existence of fractures and faults, Ks differs largely from the surrounding solid hard rock \citep{Lachassagne2021}. 

\citet{Hector2013} applied a combined geophysical survey in an unconfined weathered hard rock aquifer. Due to the spatial heterogeneity of the weathered hard rock, the prism formula was used for forward calculation of $\Delta h$ to $\Delta g$. AG is successfully used to detect the seasonal GWSC, which was measured as a seasonal water table variation of 4m. Information on the spatial heterogeneity of the underground caused by local hydrological effects was provided with magnetic resonance sounding and resistivity data, which significantly improved the fit of the measured and modeled gravity signal. In contrast, \citet{Gntner2017} assessed the performance of a long-term SG measurement for hydrological monitoring, without considering the heterogeneity of the weathered hard rock aquifer, but nonetheless presented a good fit between measured gravity residuals and estimated gravity based on hydrological fluxes. This is due to the larger area covered, a radius of 200m instead of 100m in the study of \citet{Hector2013}. Thus, local disparities are covered in the averaged signal. Another reason can be found in the scaling factors of evaporation and runoff in the water balance \citep{Gntner2017}, which might imitate the dynamics caused by heterogeneity in weathered hard rock. 

\citet{Imanishi2013} observed the dripping inside an underground tunnel in weathered rock through preferential paths, as the dripping is not uniformly distributed. The ongoing linear increase of gravity measured with an underground gravimeter after the end of dripping suggests spatial variability in water fluxes in the way that in some areas water is still flowing downwards, even after the dripping in the tunnel has stopped.

A detailed investigation of the gravity signal caused by hydrogeological conditions in weathered hard rock was conducted by \citet{Hokkanen2006}. The effect of i) surface water and ii) soil water, which perches above the bedrock due to the lower Ks of the hard rock (figure \ref{fig4}), was calculated as two different 3D G-models. The 3D model of the fractured hard rock was built based on six fractures, which were converted into triangle meshes. The models convert water storage changes into $\Delta g$ at the ground surface and above the bedrock, with the approach of \citet{Barnett1976}. The effect of $\Delta h$ on $\Delta g$ was larger for the surface water layer ($\Delta h$/ $\Delta g$ = 10mm/0.07 µGal) than for the perched soil water ($\Delta h$/ $\Delta g$ = 10mm/0.02 µGal), which is caused by the change in density that is lower in the groundwater layer (300 kg/m$^3$) than in the surface water layer (1000 kg/m$^3$), due to the porosity of the weathered hard rock. Both G-models calculated a higher $\Delta g$ than predicted by the Bouguer approximation. 

The gravity variation caused by water flow in the fractures of the bedrock above the groundwater table (in the UZ) was represented by five 3D G-models, which were refined in \citet{Hokkanen2007} with varying fracture openings of 1-5 mm. The larger the fracture opening, the larger the resulting gravity change. The spatial variation of the porosity in the weathered rock at the horizontal extent was represented by quadratic blocks of varying sizes (2m, 5m, 10m, and 20m) and porosity (0-2\%), and the resulting gravity variation was calculated. Larger blocks and porosity caused a larger increase in gravity. Long-term effects on the gravity signal are caused by GWSC within the bedrock of varying porosity and hydraulic conductivity \citep{Hokkanen2006}, whereas immediate effects are caused by fracture water \citep{Hokkanen2006}. 

The downward water movement through a fracture in the hill above an SG caused a decrease of gravity of 2 nm/s$^2$ within a 4-hour irrigation experiment, during which 20 m$^3$  of water was applied at the top of the hill \citep{Kroner2006}. \citet{Murty2002} investigated both secondary porosity (weathered parts and fractures) and primary porosity of the rock with gravimetry, which causes different densities of the underground. Both densities are included in the Bouguer model to calculate $\Delta g$. With a planar gravity survey, different geological materials, due to their varying primary porosity, could be identified. A survey along two traverses and its spectral analysis provided information on the depth of weathering and the underlying fractured zone. These physical properties were directly linked to groundwater availability \citep{Murty2002}. These studies illustrate the heterogeneity of weathered hard rock aquifers and the resulting limits for use of gravimetry, e.g. \citep{Hokkanen2006} stress the need for a sophisticated fracture model to be able to determine the distribution of water and the resulting change in gravity precisely. 

\subsection{Influence of the unsaturated zone on the gravity signal}
Mass changes in the UZ due to evapotranspiration, infiltration, and interflow are often less than in the SZ, thus the resulting gravity signal is low. Only very deep UZs, in which the larger volume can cause higher mass changes, the gravity signal from UZ exceeds that of SZ. This is also caused by the vicinity of the UZ to the sensor, even though the total mass change in UZ might be significantly lower. Table \ref{table3} gives an impression of reported gravity changes caused by UZ processes. 

\subsubsection{Gravity signal caused by infiltration}
The gravity signal caused by infiltration of precipitation into the UZ is in the range of 0.1 nm/s$^2$ \citep{Chen2020} up to approx. 90 nm/s$^2$ \citep{Creutzfeldt2010b}, depending on the quantity of SMC and the depth of the UZ. \citet{Chen2020} and \citet{VanCamp2006} measure SMC to quantify its error propagation in $\Delta g$, when estimating $Sy$. \citet{Creutzfeldt2010b} improve the fit of the measured and estimated $\Delta g$ time series based on local hydrology. The study shows that including high-accuracy measurements of rainfall, actual evapotranspiration (ETa), and SMC, using a lysimeter, increased the correspondence of the measured $\Delta g$ and the inverted $\Delta g$. The RMSE was reduced from 57 nm/s$^2$ to 6.2 nm/s$^2$, and thus 97\% of the SG residuals could be explained by the combined approach of lysimeter and H-model \citep{Creutzfeldt2010b}. Similarly, \citet{Krause2009} showed that 77\% of the gravity residuals could be explained by SMC, which was modelled for the catchment in which the SG was located. As the SG was located at a lower position in the catchment, the SMC influences the gravity residuals in a reversed manner. It is pointed out that the interception of the surrounding forest and litter layer impeded infiltration of precipitation lower than 4mm, which can be detected in the $\Delta g$ time series. Both studies stress the importance of including mass changes in UZ when interpreting $\Delta g$ signal.
\begin{table*}[htbp]
\caption{Overview of rainfall or changes in the volumetric water content (VWC) in the UZ and the resulting measured signal.}
\begin{tabular}{l l c c c p{3cm} l l}
\tophline
\textbf{Rainfall} & SMC &
$\Delta g$ & $\Delta g$ &
Depth UZ & Lithology & Device & Study \\
& & [nm s$^{-2}$] & [$\mu$Gal] & [cm] & & & \\
\middlehline

1 mm & -- & 0.305 & 0.0305 & -- & -- & -- & \citep{Longuevergne2009} \\

64 mm/day & 7 \% VWC & 6 & 0.6 & 270 & Sand & SG & \citep{Abe2006} \\

40 mm/h & -- & 5 & 0.5 & 270 & Sand & SG & \citep{Abe2006} \\

60 mm & -- & 29 & 2.9 & 110 & Loamy sand,\newline weathered gneiss & SG & \citep{Creutzfeldt2010a} \\

9 mm/h & 7.5 & 17 & 1.7 & 200 & Silty loam & SG & \citep{Krause2009} \\

25 mm/h & -- & 15 & 1.5 & 350 & Loam & SG & \citep{Longuevergne2009} \\

\middlehline
\multicolumn{8}{l}{\textbf{Irrigation experiments}} \\
\middlehline

10 days $\times$ 86.2 mm/day & 10 & 150 & 15 & 250--300 & Sand, gravel & RG & \citep{Christiansen2011a} \\

$\sim$1 mm/h & 20 & 7.5 & -- & 50 & Soil,\newline weathered hard rock & SG 30m bgl & \citep{Imanishi2013} \\

\bottomhline
\end{tabular}
\belowtable{}
\label{table3}
\end{table*}

An empirical equation estimates the gravitational signal caused by infiltration of rainfall and water losses through ET in the UZ for time $t > t_r$
\[
\Delta g_r(t) = 2 \pi G \rho_w \, r(t_r) \, f(t)
\tag{5.3}
\]
With $\Delta g_r$is the gravity effect of the rainfall, $r(tr)$ is the rainfall at time tr. The function 
\[
f(t)=\left(1-e^{-\frac{t-t_r}{\tau_1}}\right)
e^{-\frac{t-t_r}{\tau_2}}
\tag{5.4}
\]
with $\tau 1$, $\tau 2$ being recharge and discharge time parameters, describes the increase and decrease of water mass due to infiltration and ET \citep{Crossley1998a, Harnisch2006}. \citet{Pendiuk2020} used this approach to calculate the $\Delta g$ signal for the UZ and fitted their objective functions to determine the parameters $\tau_1$, $\tau_2$, assuming that $\tau_1$ is only dependent on the field capacity and thus constant. $\tau_2$ is assumed to be dependent on the air temperature $\tau_2 = a / (T(t))$, with $a$ being a fitting factor and T(t) being the mean air temperature per hour. The correlation between the simulated and the observed gravity series, over the course of 1 year, is 96\%. Thus, the G-model could explain the measured gravity to a high degree. The resulting $\Delta g$ from the UZ is mostly influenced by the short-term processes, in contrast to the signal from groundwater, which shows a seasonal pattern \citep{Pendiuk2020}. This is similar to a study from Indonesia, in which 80\% of the short-term gravity change ($\approx 0.6\,\mu\mathrm{Gal}$) was mainly caused by intense rainfall (64mm/day) and the resulting SMC \citep{Abe2006}.
A study at an underground SG in Vienna showed that the SMC in the UZ, with a depth of 4.1 m above the SG, dominates over the GWSC, which is approximately 9m below the SG \citep{Mikolaj2015}. Similarly, the seasonal peak-to-peak amplitude of the gravity signal induced by SMC in a study of an underground SG \citep{Longuevergne2009} was larger (20 nm/s$^2$ ; 2.0 µGal) than the peak-to-peak amplitude of the groundwater induced signal (10 nm/s$^2$; 1.0 µGal) in most years. Exceptionally wet years in this study caused a much larger gravity signal (peak-to-peak amplitude of 50 nm/s$^2$; 5 µGal), with the time series of the SMC being constant throughout the period of 6 years. In the same study, an admittance of -0.305 nm/s$^2$ per 1 mm water layer (above the gravimeter) was calculated \citep{Longuevergne2009}.
\subsubsection{Irrigation experiments}
These findings indicate that even though the SMC often causes a lower gravity signal, it’s effect should be considered in hydro-gravimetric studies. In the studies of \citet{Christiansen2011a, Reich2021}, the signal is increased by applying an irrigation experiment and deducing information about UZ characteristics from the measured $\Delta g$. 

With a 6 h irrigation experiment, during which 64.4 mm of water was sprinkled on an area of 15m x 15m, preferential flow and water table advancement could be identified as predominant infiltration processes, whereas water ponding upon an impermeable soil layer (perched soilwater) could be excluded. This was possible due to the deployment of a ‘hydro-gravimetric model’, which was run with synthetic data for five different scenarios of UZ conditions (macropores, preferential flow, waterfront advancement, bypass flow, perched water). The model output is a unique gravity response over time for each scenario. This response was compared to the measured gravity response during the irrigation experiment, and the best fit indicated the predominant condition. The gravity response caused by irrigation increased from 0 to 29 nm/s$^2$, which could be measured with the used SG (igrav 006) in this experiment \citep{Reich2021}. 

When gravimeters with lower accuracy are applied, longer experiments with a larger total amount of sprinkled water can be performed. \citet{Christiansen2011a} conducted a 14-day drip irrigation experiment, during which 86.2 mm/d were irrigated on an area of 10.33m • 10.33m, to successfully calibrate an UZ model. The gravity response was measured with a RG (Scintrex CG-5) in a star network with a reference station. A UZ model was set up for simulating the infiltration experiment, and the SMC in this H-model was converted to $\Delta g$. The H-model was calibrated by minimizing the objective function, which is the difference between observed and simulated gravity response. In doing so, the uncertainty of the saturated water content ($\theta s$) could be reduced, and the saturated Ks and the parameter n of the water retention model of van Genuchten (Van Genuchten, 1980) could be constrained. The lower sensitivity, resulting from the reduced gravity response in the deeper soil layers are quantified as follows: a total $\Delta g$ of 150 nm/s$^2$, of which 50 nm/s$^2$ could be attributed to the SMC in the upper meter, another 50 nm/s$^2$ to the following 1.5 m and the residual 50 nm/s$^2$ due to SMC in the soil interval of 2.5 m -- 5 m below \citep{Christiansen2011a}. 

\citet{Gntner2017} calculates the potential gravity signal caused by infiltration into the soil with the prism approximation \citep{Nagy1966}. The infiltration is simulated with Hydrus 1D and a synthetic data set of constant 24 mm/d precipitation for 15 days, followed by 15 days of evapotranspiration. The results stress the impact of an impermeable platform or building shielding the gravimeter, which prevents infiltration close to the gravimeter. The so-called ‘umbrella’ or ‘shelter’ effect causes an increase in the gravity signal even after the end of the rainfall and at the beginning of the evapotranspiration process, whereas the field set-up without a larger impervious platform causes a linear decline of the gravity residuals \citep{Gntner2017}. This effect hinders the use of gravity for estimating hydrological fluxes, as the instrument is shielded from fast WSC occurring near the surface in its immediate vicinity \citep{Hector2015}.

\subsection{Hydro-gravimetry and modelling}
Research on the combination of hydrological modelling and gravity measurements focus on either explaining residuals of gravity data \citep{Hasan2006, Lampitelli2010, Weise2018}, or calibration of the H-model by use of gravity data \citep{Christiansen2011a, Hasan2008, Kennedy2016, Kennedy2022, Kroner2006, Pivetta2021, Piccolroaz2015} or both \citep{Naujoks2010}. Changes in gravity provide a direct measure of storage change, which is a key variable in groundwater flow modeling. This, and the fact that it can characterize an entire system \citep{Kennedy2016} makes gravity data perfectly suitable for H-model calibration, as H-models provide gravity values directly \citep{Kennedy2023}. Thus, hydrogeological \citep{Kennedy2022}, hydrological \citep{Naujoks2010}, integrated hydrological and hydrogeological \citep{Hasan2008} and UZ models are calibrated with gravity measurements of varying setups and duration \citep{Christiansen2011a, Kennedy2016}. \citet{Weise2018} further developed the H-model of \citet{Naujoks2010} by including groundwater changes, snow impact and the variations in the roof above the building where the gravimeter is located. Their comparison of newly corrected gravity residuals showed a better agreement with GRACE data for the area than that in the former original work of \citet{Naujoks2010}, emphasizing the role of groundwater when calculating gravity residuals.  

\begin{figure}[t]
\includegraphics[width=12cm]{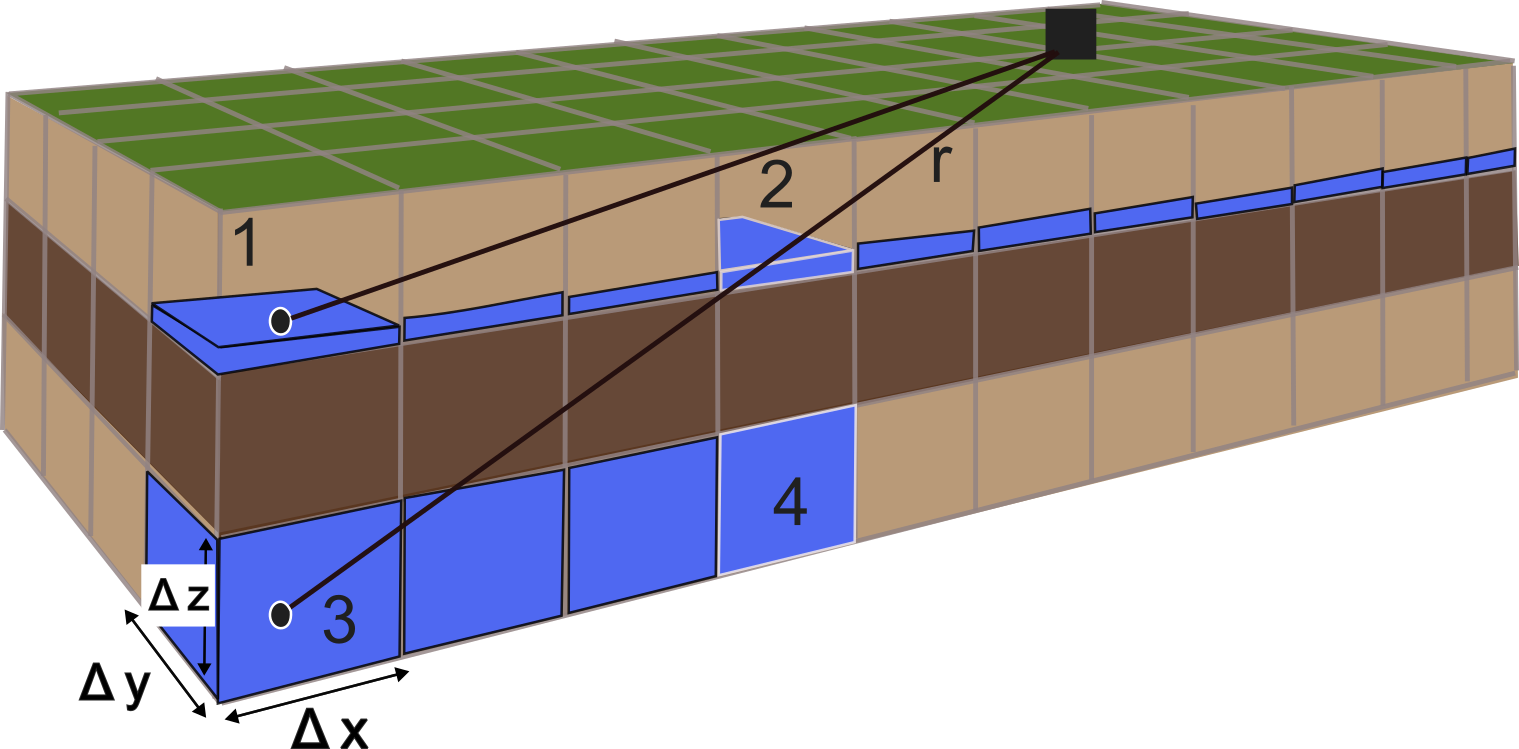}
\caption{Geometry of prismatic cells in a hydrogeological model relative to the sensor: 1) mass change in an unconfined cell (point mass formula), 2) nearby mass change in an unconfined cell (prism formula), 3) distant mass change in a confined cell overlain by a less permable layer (point mass formula), 4) nearby mass change in a confined cell (prism formula) (adapted from: \protect\citet{Kennedy2023})}
\label{fig5}
\end{figure}

A recently developed software called \textit{Heavy} for calibrating H-models with gravity field measurements implements the method from \citep{Leirio2009} for the hydrogeological software Modflow. Heavy calculates the gravity for each cubic cell in the H-model, both confined and unconfined ones, and reports the change of gravity for each layer and the cumulative change (figure \ref{fig5}). Due to the lower benefit of implementing the MacMillan formula for medium distances, the prism formula is applied for cells fulfilling the criteria f²<81. For the remaining cells, the point-mass formula is applied. The effects of surface water bodies (lakes and streams) and the UZ are neglected \citep{Kennedy2023}. Similarly GravCHAW couples MODFLOW with the Gravi4GW Software to assimilate time-lapse gravimetry and hydraulic head observations for groundwater model calibration, but it additionally adds parameter estimation and uncertainty analysis to the forward calculation. As Gravi4GW explicitly accounts for topographic relief, it is particularly well-suited for catchments with pronounced topography \citep{Mohammadi2026GravCHAW}.  

A study from Kennedy et al., 2016 investigating a managed aquifer recharge facility using a combination of gravity measurements and modelling revealed that calibration with gravity is superior compared to groundwater level data. The model combined saturated (Modflow-NWT) and UZ flow (Modflow-UZF). Gravity captured processes in the UZ better than groundwater level data and is more sensitive to events in proximity in this study. Whereas groundwater level data is impacted by all processes in the catchment, e.g., natural groundwater recharge, pumping, and irrigation, the use of gravity data enabled the use of a simpler H-model for the area of the recharge facility, without considering the whole catchment \citep{Kennedy2016}. 

\citet{Christiansen2011a} successfully calibrated an UZ model (MIKE SHE) by constraining the van Genuchten parameter n and Ks for a single-layered soil by measuring gravity during an irrigation experiment. A different approach using SG time series divides the gravity residuals into short-term and long-term variations, with the fast (hourly) variations being caused by precipitation and the slow (daily) variations by subsurface redistribution. The fast signal is used to calibrate the water-balance-based model to calculate recharge. The recharge is used as input into a hillslope-storage Boussinesq model, which is then used to estimate the slow gravity variations. The gravity effect from both H-models can explain 80\% of the variance in the measured gravity \citep{Hasan2008}.
The above methods are suitable for finite-element meshes with prismatic cells, whereas the approach described by \citet{Barnett1976, Petrovic1996} can be used to compute the gravitational effect of a triangulated mesh in finite-volume models. 
\section{Limitations of gravimetry in hydrogeology}
\subsection{Accuracy and duration of measurement}
Device-specific limitations arise from a low signal-to-noise ratio, where the device's accuracy is lower than the hydrological signal. The change in gravity caused by small storage variations, e.g., SMC, is smaller than the error of the gravimeter (FG5) \citep{Chen2002}. Nonetheless, information on SMC must be considered to determine GWSC accurately \citep{Creutzfeldt2010a, Krause2009}. So far, only the high accuracy and long-term measurements of SG can provide information about processes in the UZ \citep{Longuevergne2009, Mikolaj2015,Pendiuk2020, Reich2021}. Besides other challenges, SGs are often set up in buildings, which causes the disadvantage of the umbrella effect and thus influences the natural infiltration in the vicinity of the SG \citep{Gntner2017}. \citet{Pivetta2021} describes a strong drift of the gPhone, which could not be quantified with an AG, which limited the use of gravity data to short-term, extreme events. 

The correction required to resolve the measured gravity signal for the hydrological signal introduces additional uncertainty \citep{Mikolaj2019}. By comparing global models simulating the signals to be corrected, \citet{Mikolaj2019} determines an RMSE of 0.6 nm/s$^2$ for tidal corrections, 1.7 nm/s$^2$ for nontidal ocean loading, 2.6 nm/s$^2$ for atmospheric corrections, and 3.8 nm/s$^2$ for large-scale hydrological effects. 

Published investigations of the porous aquifer, applying RG or AG, focused on areas with GWSC larger than 1 m \citep{Chen2020, Christiansen2011c, Hsiao2021, Pool1995, Pool2008}, which causes $\Delta g$ = 126 nm/s$^2$ (12.6 µGal) considering the Bouguer approximation (eq. 4.1) and porosity of 30\%. When the duration of measurement of the final, averaged value of gravity is too long, changes in soil moisture or groundwater during this time induce errors \citep{Chen2020}, especially in highly dynamic systems such as karst aquifers. 

\subsection{Conditions in the underground}

If the degree of heterogeneity in the subsurface is too high, which is often the case in weathered hard rock and karst aquifers, or confined parts in porous aquifers \citep{Pool2008}, additional information for converting $\Delta h$ into $\Delta g$ is required, and the Bouguer approximation is not applicable \citep{Hector2013, Imanishi2013, Kroner2006}. A comparison with a suitable forward G-model is recommended in such cases, as demonstrated by \citep{Hokkanen2007}. In fact, all conditions impacting the gravity signal have to be known and separated to gain information on the GWSC. 

\subsection{Planar coverage}
The gravity signal resulting from different vertical distances between the sensor and the cylinder (h) was calculated by \citep{Gntner2017}. For a homogeneous cylinder with a height of 0.1m and a range of h between 0.25 m to 5 m, the signal reached 41.93 nm/s$^2$ at radii larger than 200m. Including an impervious platform, on which the sensor is placed, reduced the gravity signal by half, when h was set to the smallest distance of 0.25 m \citep{Gntner2017}. Thus, the conversion of $\Delta h$ into $\Delta g$ significantly depends on topography, depth to groundwater, and temporal changes in water distribution in the soil. 

\begin{figure}[t]
\includegraphics[width=8.3cm]{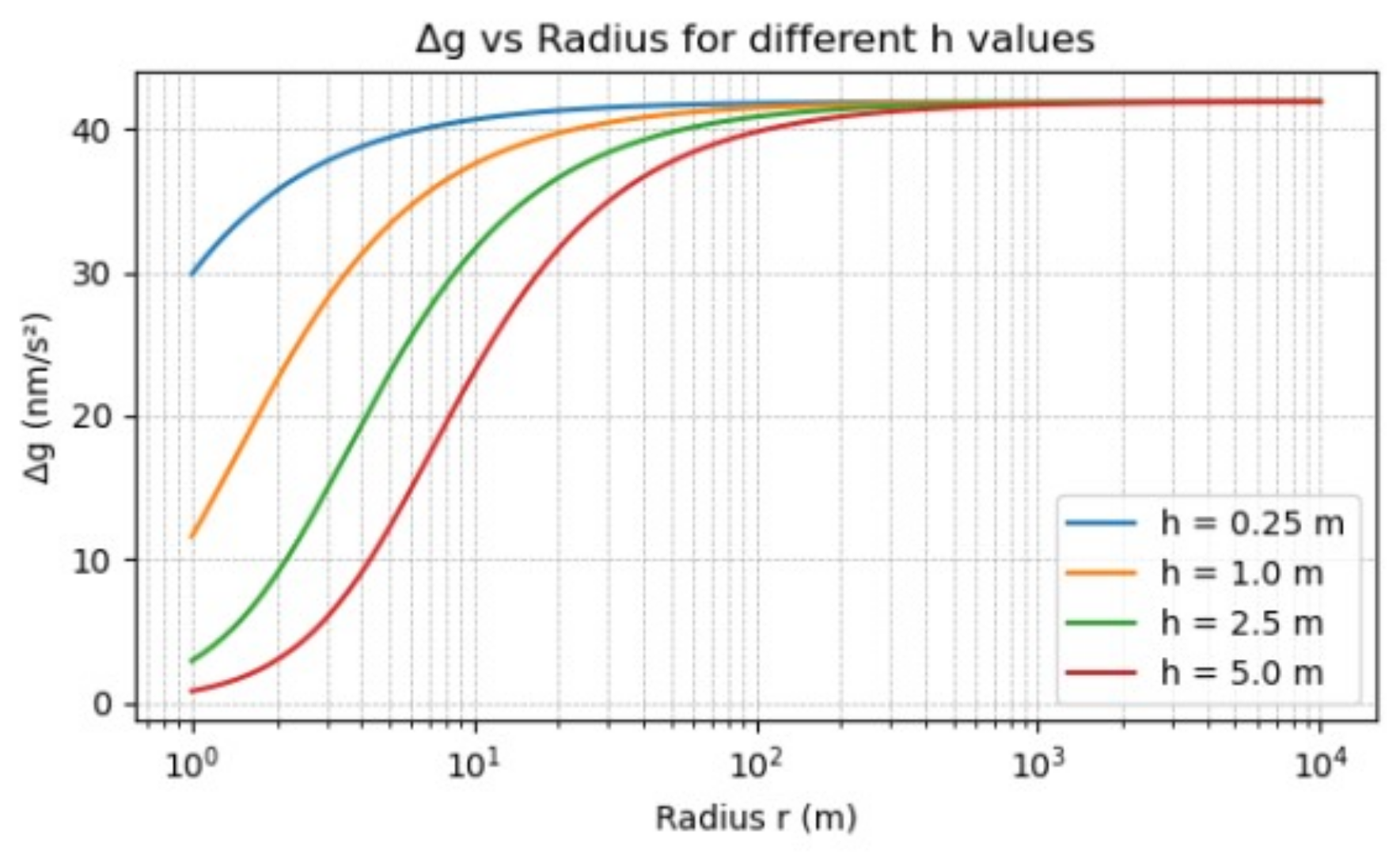}
\caption{Gravity effect of a homogenous cylinder with a thickness of 10 cm, a density of 1000 kg/m$^3$ and varying radii on a sensor placed at different heights (0.25 to 5 m) above the cylinder (adapted from \protect\citet{Gntner2017}, published under CC BY 3.0)}
\label{fig6}
\end{figure}

According to the Bouguer approximation, if the radius of the homogeneous cylinder is sufficiently large, the gravimetric signal is only dependent on the thickness and density of the layer and independent of the distance between the layer and the sensor \citep{Gntner2017}. This is true for a radius larger than 200 m for sensor heights smaller than 5m (figure \ref{fig6}). A further increase in radius does not cause an increase in $\Delta g$. The SG (iGrav) needs to be placed on solid ground, which is constructed from mostly impervious concrete. This platform reduces the gravity signal resulting from the infiltrating water when the sensor is too close to the cylinder \citep{Gntner2017}. Thus, the distance between the sensor and the investigated water table of at least 2.5 m is required. The platform should be as small and permeable as possible. On the other hand, a larger distance between sensor and layer causes a smaller signal in the vicinity of the sensor, if the lateral extent of the layer of mass change is <100m)

\citep{Hector2013} calculated the footprint of the gravimeter, which represents the area that contributes most significantly to the observed gravity signal, with a 3D G-model based on the prism formula \citep{Leirio2009}. This footprint was determined by comparing the gravity effect of a 1m thick water layer within concentric disks of increasing radius to the effect of the same layer distributed in a much larger disk with a radius of 1000 m. The results indicate that 90\% of the gravity signal produced by the shallow groundwater layer originates from relatively small radii \citep{Gntner2017}. Specifically, for a typical water table depth of 5 m below ground level (bgl), 92\% of the gravity signal accumulated within a 1000 m radius is already captured within the inner 100 meters. This confirms the 100 m radius as a suitable and effective footprint for homogeneous underground and a sensor-to-layer distance of 5m. 
\section{Future directions}
The development of devices is aiming at improved accuracy and reproducibility, which affects the potential application of gravimetry for hydrological water mass changes and further subsurface monitoring.
\subsection{Instruments}
Over the past two decades, research in atomic gravimetry has seen significant advancements with the development of various prototypes, achieving resolutions from a few $\mu$gals to record lows in the sub-$\mu$gal range, as demonstrated by projects like GAIN \citep{Freier2016GAIN}. This progress has led to the emergence of multiple commercial Atom Gravimeters, which have been rigorously tested and compared in numerous campaigns alongside other devices (Appendix \ref{sec:listGravi}). The leading models achieve long-term reproducibility below 10 $\mu$gal. 
However, this reproducibility is generally affected by several factors, with a notable influence from the spatial convolution between the atomic matter waves and the light field responsible for atomic interference. To achieve long-term reproducibility at the $\mu$gal level, it is crucial to further reduce these systematic biases. Enhancing the spatial mode quality of the light field used in interferometry is essential, but also utilizing ultracold or even quantum degenerate atomic ensembles, rather than merely laser-cooled ones, may mitigate biases arising from kinematic effects within the interferometric light field, as suggested by studies such as \citep{Karcher2018UltracoldSourcesInterferometers}. 
Recent advancements in trapping techniques using atom chips or optical dipole traps have enabled the rapid and reliable generation of these ultracold ensembles and even quantum degenerate gases \citep{Abend2016AtomChipGravimeter}. These techniques are increasingly being integrated with atomic gravimeters, as discussed in e.g. \citet{Heine2020TransportableQuantumGravimeter} and \cite{Karcher2018UltracoldSourcesInterferometers}. Atom chip-based sources for quantum degenerate matter have been successfully demonstrated in space through missions like MAIUS-1 \citep{Becker2018SpaceBorneBEC}, paving the way for more compact setups, as explored in studies as explored by \citet{Abend2016AtomChipGravimeter}. Similar developments are underway for dipole traps, as noted by \citet{Anton2025INTENTAS}. Additionally, considerable effort is focused on the miniaturization of gravimeter components to decrease their size, weight, and power consumption, thereby enhancing their robustness and transportability. Looking ahead, atom interferometers with extended baselines and capabilities for large momentum transfer could potentially also complement superconducting gravimeters. By fusing short-term stability with absolute measurements, they may serve as gravimetric references, helping to validate the long-term reproducibility of gravimeters to the precision required for applications such as hydrology, as highlighted in \citet{Schilling2020VLBAIGrav}.
\subsection{Potential future applications}
AQG are having an impact in several industries, ranging from aerospace and terrestrial navigation, underground resource exploration, land- and water monitoring \citep{Oh2024}. Gravimeters deployed on aircrafts can monitor inaccessible areas and survey critical infrastructure \citep{DadrassJavan2025, KUMAR2022, bidel2023}. The ability to leverage the natural properties of an atom results in a device with potential for higher stability, accuracy, and $\mu$Gal level resolution. Advancements in atom interferometry are enabling portable, high-resolution instruments capable of detecting minute variations in gravitational fields \citep{bongs2019, degen2017}, and miniaturization of the atomic trapping and laser cooling can lead to a reduction in size and power requirements \citep{narducci2022}. A fundamental enabling technology for compact and cost-effective cold-atom gravimeters lies in the precise stabilisation of laser frequencies. To achieve accurate preparation, manipulation, and detection of ultracold atoms, laser light must remain locked to a stable atomic reference, as even small frequency drifts introduce significant measurement errors. 

Groundwater monitoring is a critical component of sustainable water management, yet traditional hydrogeological methods face limitations in spatial coverage, temporal resolution, and subsurface mass detection. Current practices must evolve to be able to provide high-resolution data \citep{rau2020}. RGs (spring, MEMS, or SGs) have been able to detect subtle variations in gravitational fields. However, these monitoring systems require frequent calibration to a reference point, drift over time, and are logistically challenging to deploy in the field \citep{Shettell2024}, which adds uncertainty to measuring groundwater efficiently. 

Quantum gravimetry, especially with portable AQGs, is highly promising for groundwater monitoring because it combines high sensitivity (1$\mu$Gal after one hour) and stability with ease of deployment \citep{Cook1989}. They have the potential to significantly enhance the economic efficiency of gravity campaigns by reducing time and resource demands, while simultaneously delivering higher measurement accuracy. It has been demonstrated that portable gravimeters constitute a non-destructive, operationally, financially sustainable, and efficient monitoring system for measuring groundwater storage change when combined with a hydrogeological model \citep{El-Diasty2016, Cooke2021} and will therefore be an important tool in sustainable non-invasive groundwater management in the face of climate change \citep{jakeman2016}. 

The EQUIP‑G project, funded by Horizon Europe, is also demonstrating how AQG is transitioning into commercial applications and advancing from traditional RG. EQUIP‑G is deploying a network of field-ready AQG across Europe to provide absolute, drift‑free gravity reference data. Led by major research institutes like Centre National de la Recherche Scientifique (CNRS), Center for Geosciences (GFZ) or Finnish Geospatial Research Institute (FGI) and industry partners, the project is moving atom‑interferometry gravimetry from laboratory prototypes to a functional geoscience infrastructure \citep{EC_NEWTONg}. 

\conclusions
This review summarizes the latest progress in the development of gravimeters, with special focus on field deployment and their application for hydrological and hydrogeological monitoring. 

Currently developed gravimeters based on atom interferometry provide measurements of absolute gravity with high precision, which can extend applications in hydrogeology, as the devices are applicable in field conditions. Recently published software and calculation approaches to determine the gravity changes caused by GWSC additionally foster the use of gravimeters. Combining AQG with MEMS RG, which are small, cost-effective, and can be placed flexibly, potentially increases spatial and temporal resolution. 

With gravimetry, the spectrum of possibilities to monitor groundwater storage change is enhanced, providing opportunities beyond traditional methods, such as integrated estimation of $Sy$, measuring without intruding the underground, or monitoring in difficult conditions like alpine catchments. Constraining hydrogeological models with gravity measurements as an additional reference figure is promising. Considering the UZ processes within the modelling approach is necessary to resolve the measured gravity signal into its governing subsurface processes.
Groundwater resources are globally under stress, requiring sustainable management. Gravimetry can serve as a novel monitoring tool, enhancing the overall data basis and supporting more informed management decisions.

\section{Glossary}
\begin{description}
\item[Accuracy] The degree of agreement between a measured value and the true or accepted reference value.
\item [Confined aquifer] A low-permeability layer overlies the aquifer.
\item[Drift] A slow, non-stochastic change in the instrument reading over time caused by internal effects (e.g., aging of a spring) rather than actual changes in gravity.
\item[Interferometer]An optical device that uses the interference of light (or matter) waves to measure distances.
\item[Reproducibility] The closeness of agreement between independent measurement results of the same quantity obtained under changed conditions (e.g., different instruments or locations).
\item[Precision] Stability of the measurements, defined as the deviation of individual measurements with respect to their average value.
\item[Sensitivity]: The degree to which an instrument’s output varies for a specified change in the measured quantity.
\item[Stability] Change in accuracy over time; it represents the total variation in measurements of the same part measured over time.
\item[Specific yield] Volume of water per unit volume of an aquifer that can be extracted by gravity drainage.
\item[Saturated Zone] Zone below the groundwater table in which the pore space is fully saturated with water.
\item[Unconfined Aquifer] The aquifer has a connection to the atmosphere through the vadose zone. 
\item[Unsaturated Zone, Vadose Zone] Zone below Earth's surface and above the groundwater table, where part of the pore space is filled with air.
\item[Alluvial aquifer] Groundwater stored in alluvial deposits (gravel, sand, silt, or clay).
\item[Hard rock aquifer] Groundwater stored in the weathered parts of hard rock.
\end{description}

\appendix
\section{Concept of Weathered Hard Rock}
\label{sec:hardrock}
\begin{figure}[h]
\includegraphics[scale=0.9]{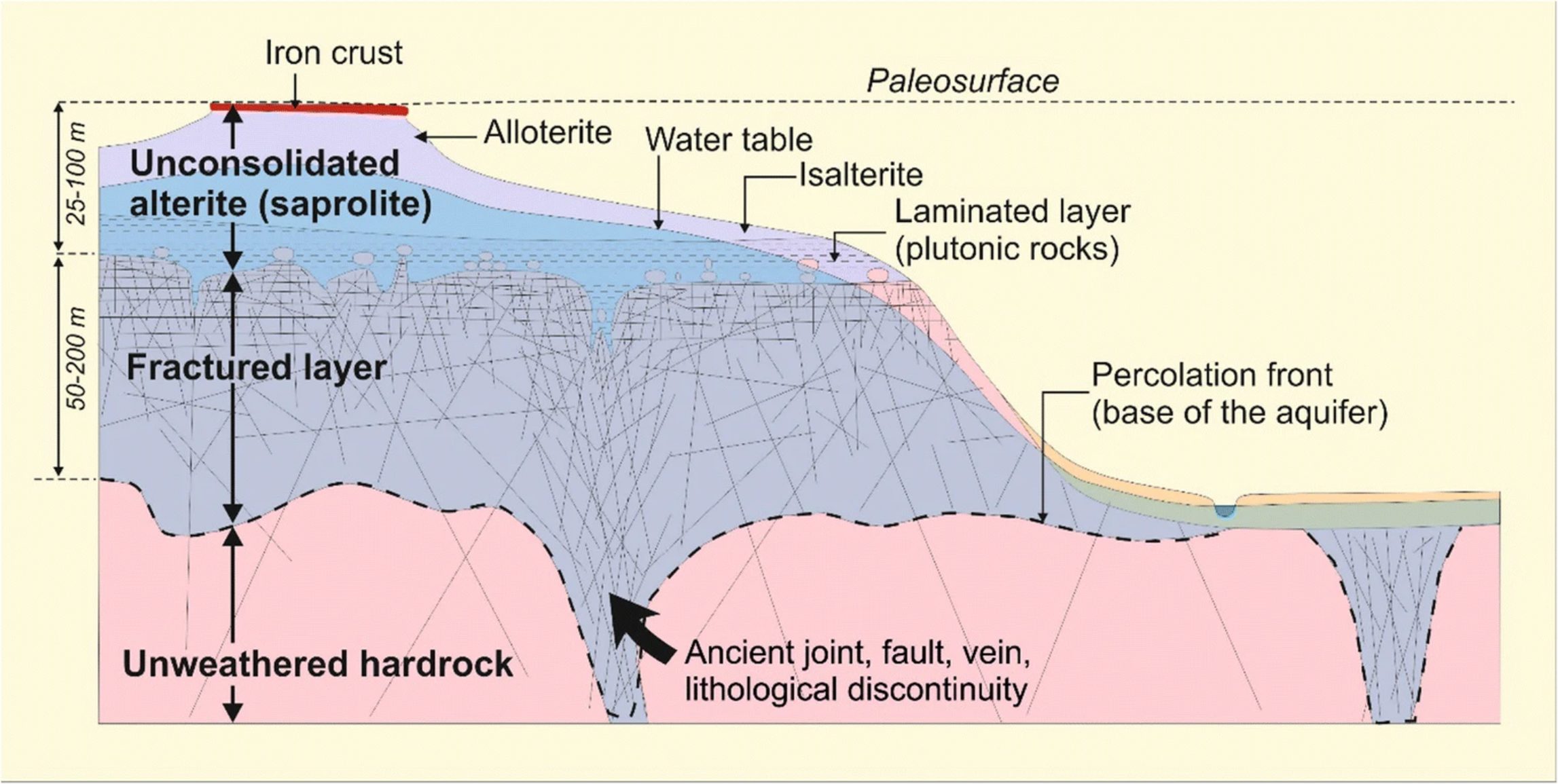}
\caption{Conceptual model of a partly eroded weathering profile on hard rocks (erosion = right part of the figure). Note that all technical terms in this figure are described and explained in section ‘Geological structure and hydrodynamic properties of weathering profiles’ in \protect\citet{Lachassagne2021}. From \protect\citet{Lachassagne2021}, published under CC BY 4.0}
\label{figsm1}
\end{figure}
\newpage

\section{MacMillan Formula}
\label{sec:mcm}
The MacMillan formula is applied for intermediate distances between the sensor and model cell \citep{MacMillan1958, Leirio2009}:
\[
\Delta g = G \rho_w Sy \, \Delta x \, \Delta y \, \Delta z \left(
-\frac{z}{d^3}
- \frac{5}{24} \cdot \frac{(\alpha x^2 + \beta y^2 + \omega z^2) z}{d^7}
+ \frac{1}{12} \cdot \frac{\omega z}{d^5}
\right)
\]

\[
\text{with} \quad
\alpha = 2 \Delta x^2 - \Delta y^2 - \Delta z^2, \quad
\beta = -\Delta x^2 + 2 \Delta y^2 - \Delta z^2, \quad
\omega = -\Delta x^2 - \Delta y^2 - 2 \Delta z^2
\]

\section{List of Commercial Atomic Gravimeters}
\label{sec:listGravi}
\begin{itemize}

\item \textbf{Exail (Muquans AQG)} \\
\url{https://www.exail.com/product/quantum-gravimeters}

\item \textbf{M Squared Lasers – Quantum Gravimeter} \\
\url{https://m2lasers.com/quantum-gravimeter.html}

\item \textbf{QuantumCTek – Cold Atomic Gravimeter} \\
\url{https://www.quantum-info.com/English/product/Quantum_Precision_Measurement/Quan/2024/1211/870.html}

\item \textbf{Lonten / Lontenoe – Cold Atomic Gravimeter} \\
\url{https://lontenoe.com/products/cold-atomic-gravimeter/}

\item \textbf{CAS Cold Atom Technology (CASCA, Wuhan)} \\
Company: \url{https://www.cascoldatom.com/en/direction.html} \\
Product page: \url{https://www.cascoldatom.com/en/more.php?lm=10&id=104}

\item \textbf{Vector Atomic} \\
\url{https://vectoratomic.com}

\item \textbf{Nomad Atomics} \\
\url{https://www.nomadatomics.com}

\item \textbf{Atomionics (Gravio)} \\
Company: \url{https://www.atomionics.com} \\
Product info: \url{https://qcve.org/atomionics}

\item \textbf{Q-CTRL – Quantum sensing / navigation} \\
\url{https://q-ctrl.com}

\item \textbf{AOSense (Mark Kasevich)} \\
\url{https://www.aosense.com}

\end{itemize}

\noappendix

\appendixfigures

\appendixtables

\authorcontribution{AP, BL, CF, ER, EV, GW, IS, JA, MMD, PBG, SC, TL, and VS contributed to the literature review and interpretation. IS conceptualized the review and coordinated the work. VS initiated the project. All authors reviewed and approved the manuscript.}

\competinginterests{The authors declare that they have no known competing financial interests or personal relationships that could have appeared to influence the work reported in this paper.}

\begin{acknowledgements}
We gratefully acknowledge Sébastien Merlet (Fig. 1a) and Exail (Fig. 1c) for granting permission to use their photos in this manuscript. ChatGPT (OpenAI) was used to assist with the conversion of the manuscript from Microsoft Word to LaTeX.
\end{acknowledgements}

\financialsupport{This research has been supported by the German Federal Ministry of Research, Technology and Space (BMFTR), grant no. 13N17097.
}

\bibliography{rsc.bib}

\begin{thebibliography}{141}
\providecommand{\natexlab}[1]{#1}
\providecommand{\url}[1]{\texttt{#1}}
\providecommand{\urlprefix}{}
\expandafter\ifx\csname urlstyle\endcsname\relax
  \providecommand{\doi}[1]{https://doi.org/\discretionary{}{}{}#1}\else
  \providecommand{\doi}{https://doi.org/\discretionary{}{}{}\begingroup \urlstyle{rm}\Url}\fi

\bibitem[{Abe et~al.(2006)Abe, Takemoto, Fukuda, Higashi, Imanishi, Iwano, Ogasawara, Kobayashi, Dwipa, and Kusuma}]{Abe2006}
Abe, M., Takemoto, S., Fukuda, Y., Higashi, T., Imanishi, Y., Iwano, S., Ogasawara, S., Kobayashi, Y., Dwipa, S., and Kusuma, D.~S.: Hydrological effects on the superconducting gravimeter observation in Bandung, Journal of Geodynamics, 41, 288--295, \doi{10.1016/j.jog.2005.08.030}, 2006.

\bibitem[{Abend et~al.(2016)Abend, Gebbe, Gersemann, Ahlers, M{\"u}ntinga, Giese, Gaaloul, Schubert, L{\"a}mmerzahl, Ertmer, Schleich, and Rasel}]{Abend2016AtomChipGravimeter}
Abend, S., Gebbe, M., Gersemann, M., Ahlers, H., M{\"u}ntinga, H., Giese, E., Gaaloul, N., Schubert, C., L{\"a}mmerzahl, C., Ertmer, W., Schleich, W.~P., and Rasel, E.~M.: Atom-Chip Fountain Gravimeter, Physical Review Letters, 117, 203\,003, \doi{10.1103/PhysRevLett.117.203003}, 2016.

\bibitem[{Anton et~al.(2025)Anton, Br{\"o}ckel, Derr, Fieguth, Franzke, G{\"a}rtner, Giese, Haase, Hamann, Heidt, Kanthak, Klempt, Kruse, Krutzik, Kubitza, Lotz, M{\"u}ller, Pahl, Rasel, Schiemangk, Schleich, Schwertfeger, Wicht, and W{\"o}rner}]{Anton2025INTENTAS}
Anton, O., Br{\"o}ckel, I., Derr, D., Fieguth, A., Franzke, M., G{\"a}rtner, M., Giese, E., Haase, J.~S., Hamann, J., Heidt, A., Kanthak, S., Klempt, C., Kruse, J., Krutzik, M., Kubitza, S., Lotz, C., M{\"u}ller, K., Pahl, J., Rasel, E.~M., Schiemangk, M., Schleich, W.~P., Schwertfeger, S., Wicht, A., and W{\"o}rner, L.: INTENTAS -- an entanglement-enhanced atomic sensor for microgravity, EPJ Quantum Technology, 12, 26, \doi{10.1140/epjqt/s40507-025-00330-9}, 2025.

\bibitem[{Antoni-Micollier et~al.(2022)Antoni-Micollier, Carbone, M{\'e}noret, Lautier-Gaud, King, Greco, Messina, Contrafatto, and Desruelle}]{Micollier2020}
Antoni-Micollier, L., Carbone, D., M{\'e}noret, V., Lautier-Gaud, J., King, T., Greco, F., Messina, A., Contrafatto, D., and Desruelle, B.: Detecting Volcano-Related Underground Mass Changes With a Quantum Gravimeter, Geophysical Research Letters, 49, e2022GL097\,814, \doi{https://doi.org/10.1029/2022GL097814}, e2022GL097814 2022GL097814, 2022.

\bibitem[{Arnal et~al.(2023)Arnal, Faure, Carbone, Greco, M{\'e}noret, Rosenbusch, and Desruelle}]{Arnal2023_QuantumGravitySensors}
Arnal, M., Faure, L., Carbone, D., Greco, F., M{\'e}noret, V., Rosenbusch, P., and Desruelle, B.: Recent advances in quantum gravity sensors, in: XXVIII General Assembly of the International Union of Geodesy and Geophysics (IUGG 2023, Berlin), \doi{10.57757/IUGG23-2288}, 2023.

\bibitem[{Arnoux et~al.(2020)Arnoux, Halloran, Berdat, and Hunkeler}]{Arnoux2020}
Arnoux, M., Halloran, L.~J., Berdat, E., and Hunkeler, D.: Characterizing seasonal groundwater storage in alpine catchments using time-lapse gravimetry, water stable isotopes and water balance methods, Hydrological Processes, 34, 4319--4333, \doi{10.1002/hyp.13884}, 2020.

\bibitem[{Barnett(1976)}]{Barnett1976}
Barnett, C.~T.: Theoretical Modeling of the Magnetic Gravitational Fields of an Arbitrarily Three-Dimensional Body, Tech. rep., \urlprefix\url{http://library.seg.org/}, 1976.

\bibitem[{Basu et~al.(2022)Basu, Meter, Byrnes, Cappellen, Brouwer, Jacobsen, Jarsj{\"o}, Rudolph, Cunha, Nelson, Bhattacharya, Destouni, and Olsen}]{Basu2022}
Basu, N.~B., Meter, K. J.~V., Byrnes, D.~K., Cappellen, P.~V., Brouwer, R., Jacobsen, B.~H., Jarsj{\"o}, J., Rudolph, D.~L., Cunha, M.~C., Nelson, N., Bhattacharya, R., Destouni, G., and Olsen, S.~B.: Managing nitrogen legacies to accelerate water quality improvement, \doi{10.1038/s41561-021-00889-9}, 2022.

\bibitem[{Becker et~al.(2018)Becker, Lachmann, Seidel, Ahlers, Dinkelaker, Grosse, Hellmig, M{\"u}ntinga, Schkolnik, Wendrich et~al.}]{Becker2018SpaceBorneBEC}
Becker, D., Lachmann, M.~D., Seidel, S.~T., Ahlers, H., Dinkelaker, A.~N., Grosse, J., Hellmig, O., M{\"u}ntinga, H., Schkolnik, V., Wendrich, T., et~al.: Space-borne Bose--Einstein condensation for precision interferometry, Nature, 562, 391--395, \doi{10.1038/s41586-018-0605-1}, 2018.

\bibitem[{Bidel et~al.(2023)Bidel, Zahzam, Bresson, Blanchard, Bonnin, Bernard, Cadoret, Jensen, Forsberg, Salaun, Lucas, Lequentrec-Lalancette, Rouxel, Gabalda, Seoane, Vu, Bruinsma, and Bonvalot}]{bidel2023}
Bidel, Y., Zahzam, N., Bresson, A., Blanchard, C., Bonnin, A., Bernard, J., Cadoret, M., Jensen, T.~E., Forsberg, R., Salaun, C., Lucas, S., Lequentrec-Lalancette, M.~F., Rouxel, D., Gabalda, G., Seoane, L., Vu, D.~T., Bruinsma, S., and Bonvalot, S.: Airborne Absolute Gravimetry With a Quantum Sensor, Comparison With Classical Technologies, Journal of Geophysical Research: Solid Earth, 128, e2022JB025\,921, \doi{10.1029/2022JB025921}, 2023.

\bibitem[{Binley et~al.(2015)Binley, Hubbard, Huisman, Revil, Robinson, Singha, and Slater}]{binley2015}
Binley, A., Hubbard, S.~S., Huisman, J.~A., Revil, A., Robinson, D.~A., Singha, K., and Slater, L.~D.: The emergence of hydrogeophysics for improved understanding of subsurface processes over multiple scales, Water Resources Research, 51, 3837--3866, \doi{https://doi.org/10.1002/2015WR017016}, 2015.

\bibitem[{Blainey et~al.(2007)Blainey, Ferr{\'e}, and Cordova}]{Blainey2007}
Blainey, J.~B., Ferr{\'e}, T. P.~A., and Cordova, J.~T.: Assessing the likely value of gravity and drawdown measurements to constrain estimates of hydraulic conductivity and specific yield during unconfined aquifer testing, Water Resources Research, 43, \doi{https://doi.org/10.1029/2006WR005678}, 2007.

\bibitem[{Bongs et~al.(2019)Bongs, Holynski, Vovrosh, Bouyer, Condon, Rasel, Schubert, Schleich, and Roura}]{bongs2019}
Bongs, K., Holynski, M., Vovrosh, J., Bouyer, P., Condon, G., Rasel, E.~M., Schubert, C., Schleich, W.~P., and Roura, A.: Taking atom interferometric quantum sensors from the laboratory to real-world applications, Nature Reviews Physics, 1, 731--739, \doi{10.1038/s42254-019-0117-4}, 2019.

\bibitem[{Breili and Pettersen(2009)}]{Breili2009}
Breili, K. and Pettersen, B.~R.: Effects of surface snow cover on gravimetric observations, Journal of Geodynamics, 48, 16--22, \doi{10.1016/j.jog.2009.04.001}, 2009.

\bibitem[{Caballero et~al.(2002)Caballero, Jomelli, Chevallier, and Ribstein}]{Caballero2002}
Caballero, Y., Jomelli, V., Chevallier, P., and Ribstein, P.: Hydrological characteristics of slope deposits in high tropical mountains (Cordillera Real, Bolivia), CATENA, 47, 101--116, \doi{https://doi.org/10.1016/S0341-8162(01)00179-5}, 2002.

\bibitem[{Camp et~al.(2006)Camp, Vanclooster, Crommen, Petermans, Verbeeck, Meurers, van Dam, and Dassargues}]{VanCamp2006}
Camp, M.~V., Vanclooster, M., Crommen, O., Petermans, T., Verbeeck, K., Meurers, B., van Dam, T., and Dassargues, A.: Hydrogeological investigations at the Membach station, Belgium, and application to correct long periodic gravity variations, Journal of Geophysical Research: Solid Earth, 111, \doi{10.1029/2006JB004405}, 2006.

\bibitem[{Camp et~al.(2016)Camp, de~Viron, Pajot-M{\'e}tivier, Casenave, Watlet, Dassargues, and Vanclooster}]{VanCamp2016}
Camp, M.~V., de~Viron, O., Pajot-M{\'e}tivier, G., Casenave, F., Watlet, A., Dassargues, A., and Vanclooster, M.: Direct measurement of evapotranspiration from a forest using a superconducting gravimeter, Geophysical Research Letters, 43, 10,225--10,231, \doi{10.1002/2016GL070534}, 2016.

\bibitem[{Carbone et~al.(2020)Carbone, Antoni-Micollier, Hammond, de~Zeeuw-van Dalfsen, Rivalta, Bonadonna, Messina, Lautier-Gaud, Toland, Koymans, Anastasiou, Bramsiepe, Cannav{\`o}, Contrafatto, Frischknecht, Greco, Marocco, Middlemiss, M{\'e}noret, Noack, Passarelli, Paul, Prasad, Siligato, and Vermeulen}]{Carbone2020_NEWTONg}
Carbone, D., Antoni-Micollier, L., Hammond, G., de~Zeeuw-van Dalfsen, E., Rivalta, E., Bonadonna, C., Messina, A., Lautier-Gaud, J., Toland, K., Koymans, M., Anastasiou, K., Bramsiepe, S., Cannav{\`o}, F., Contrafatto, D., Frischknecht, C., Greco, F., Marocco, G., Middlemiss, R., M{\'e}noret, V., Noack, A., Passarelli, L., Paul, D., Prasad, A., Siligato, G., and Vermeulen, P.: The NEWTON-g Gravity Imager: Toward New Paradigms for Terrain Gravimetry, Frontiers in Earth Science, 8, 573\,396, \doi{10.3389/feart.2020.573396}, 2020.

\bibitem[{Chen et~al.(2022)Chen, Cazenave, Dahle, Llovel, Panet, Pfeffer, and Moreira}]{Chen2022}
Chen, J., Cazenave, A., Dahle, C., Llovel, W., Panet, I., Pfeffer, J., and Moreira, L.: Applications and Challenges of GRACE and GRACE Follow-On Satellite Gravimetry, \doi{10.1007/s10712-021-09685-x}, 2022.

\bibitem[{Chen et~al.(2020)Chen, Hwang, Chang, Tsai, Yeh, Cheng, Ke, and Feng}]{Chen2020}
Chen, K.~H., Hwang, C., Chang, L.~C., Tsai, J.~P., Yeh, T. C.~J., Cheng, C.~C., Ke, C.~C., and Feng, W.: Measuring Aquifer Specific Yields With Absolute Gravimetry: Result in the Choushui River Alluvial Fan and Mingchu Basin, Central Taiwan, Water Resources Research, 56, \doi{10.1029/2020WR027261}, 2020.

\bibitem[{Chen et~al.(2002)Chen, Liu, and Huang}]{Chen2002}
Chen, S.-K., Liu, C.~W., and Huang, H.-C.: Analysis of water movement in paddy rice fields (II) simulation studies, Journal of Hydrology, 268, 259--271, \doi{https://doi.org/10.1016/S0022-1694(02)00180-4}, 2002.

\bibitem[{Christiansen et~al.(2011{\natexlab{a}})Christiansen, Binning, Rosbjerg, Andersen, and Bauer-Gottwein}]{Christiansen2011c}
Christiansen, L., Binning, P.~J., Rosbjerg, D., Andersen, O.~B., and Bauer-Gottwein, P.: Using time-lapse gravity for groundwater model calibration: An application to alluvial aquifer storage, Water Resources Research, 47, \doi{10.1029/2010WR009859}, 2011{\natexlab{a}}.

\bibitem[{Christiansen et~al.(2011{\natexlab{b}})Christiansen, Haarder, Hansen, Looms, Binning, Rosbjerg, Andersen, and Bauer-Gottwein}]{Christiansen2011a}
Christiansen, L., Haarder, E.~B., Hansen, A.~B., Looms, M.~C., Binning, P.~J., Rosbjerg, D., Andersen, O.~B., and Bauer-Gottwein, P.: Calibrating Vadose Zone Models with Time-Lapse Gravity Data, Vadose Zone Journal, 10, 1034--1044, \doi{10.2136/vzj2010.0127}, 2011{\natexlab{b}}.

\bibitem[{Christiansen et~al.(2011{\natexlab{c}})Christiansen, Lund, Andersen, Binning, Rosbjerg, and Bauer-Gottwein}]{Christiansen2011b}
Christiansen, L., Lund, S., Andersen, O.~B., Binning, P.~J., Rosbjerg, D., and Bauer-Gottwein, P.: Measuring gravity change caused by water storage variations: Performance assessment under controlled conditions, Journal of Hydrology, 402, 60--70, \doi{10.1016/j.jhydrol.2011.03.004}, - test with an indoor basin<br/><br/>- 3 relative gravimeter (CG-5) and 1 absolute<br/><br/>- drifts of the relative gravimeter<br/><br/>- network measurement<br/><br/>- drift of the instrument --> larger uncertainties<br/>, 2011{\natexlab{c}}.

\bibitem[{Colombo and Rovetta(2018)}]{colombo2018}
Colombo, D. and Rovetta, D.: Coupling strategies in multiparameter geophysical joint inversion, Geophysical Journal International, 215, 1171--1184, \doi{10.1093/gji/ggy341}, 2018.

\bibitem[{Cook et~al.(1989)Cook, Walker, and Jolly}]{Cook1989}
Cook, P.~G., Walker, G.~R., and Jolly, I.~D.: Spatial Variability of Groundwater Recharge in a Semiarid Region, Tech. rep., 1989.

\bibitem[{Cooke et~al.(2021)}]{Cooke2021}
Cooke, D. et~al.: First evaluation of an absolute quantum gravimeter (AQG\#B01) for future field experiments, Geoscientific Instrumentation, Methods and Data Systems, 10, 297--305, \doi{10.5194/gi-10-297-2021}, 2021.

\bibitem[{Cordier et~al.(2026)Cordier, Leykauf, Hedges, Wrobel, Ross, MacKenzie, Freier, Wigley, and Hardman}]{Cordier2026}
Cordier, M., Leykauf, B., Hedges, S., Wrobel, A., Ross, K., MacKenzie, S., Freier, C., Wigley, P., and Hardman, K.: Quantum absolute gravimeter designed for field surveys, in: EGU General Assembly 2026, \doi{10.5194/egusphere-egu26-23076}, eGU26-23076, 2026.

\bibitem[{Creutzfeldt et~al.(2010{\natexlab{a}})Creutzfeldt, G{\"u}ntner, Thoss, Merz, and Wziontek}]{Creutzfeldt2010a}
Creutzfeldt, B., G{\"u}ntner, A., Thoss, H., Merz, B., and Wziontek, H.: Measuring the effect of local water storage changes on in situ gravity observations: Case study of the Geodetic Observatory Wettzell, Germany, Water Resources Research, 46, 1--15, \doi{10.1029/2009WR008359}, 2010{\natexlab{a}}.

\bibitem[{Creutzfeldt et~al.(2010{\natexlab{b}})Creutzfeldt, G{\"u}ntner, Wziontek, and Merz}]{Creutzfeldt2010b}
Creutzfeldt, B., G{\"u}ntner, A., Wziontek, H., and Merz, B.: Reducing local hydrology from high-precision gravity measurements: A lysimeter-based approach, Geophysical Journal International, 183, 178--187, \doi{10.1111/j.1365-246X.2010.04742.x}, 2010{\natexlab{b}}.

\bibitem[{Crossley et~al.(1999)Crossley, Hinderer, Casula, Francis, Hsu, Imanishi, Jentzsch, K{\"a}{\"a}ri{\"a}inen, Merriam, Meurers, Neumeyer, Richter, Shibuya, Sato, and van Dam}]{Crossley1999}
Crossley, D., Hinderer, J., Casula, G., Francis, O., Hsu, H.-T., Imanishi, Y., Jentzsch, G., K{\"a}{\"a}ri{\"a}inen, J., Merriam, J.~B., Meurers, B., Neumeyer, J., Richter, B., Shibuya, K., Sato, T., and van Dam, T.: Network of superconducting gravimeters benefits a number of disciplines, Eos, Transactions American Geophysical Union, 80, 121--126, \doi{10.1029/99EO00079}, 1999.

\bibitem[{Crossley et~al.(2004)Crossley, Hinderer, and Boy}]{Crossley2004}
Crossley, D., Hinderer, J., and Boy, J.~P.: Regional gravity variations in Europe from superconducting gravimeters, Journal of Geodynamics, 38, 325--342, \doi{10.1016/j.jog.2004.07.014}, 2004.

\bibitem[{Crossley et~al.(1998)Crossley, Xu, and Dam}]{Crossley1998a}
Crossley, D.~J., Xu, S., and Dam, T.~V.: Comprehensive analysis of 2 years of SG data from Table Mountain, Colorado, in: Proceedings of the 13th International Symposium on Earth Tides, pp. 659--668, Royal Observatory of Belgium, 1998.

\bibitem[{Dadrass~Javan et~al.(2025)Dadrass~Javan, Samadzadegan, Toosi, and van~der Meijde}]{DadrassJavan2025}
Dadrass~Javan, F., Samadzadegan, F., Toosi, A., and van~der Meijde, M.: Unmanned Aerial Geophysical Remote Sensing: A Systematic Review, Remote Sensing, 17, 110, \doi{10.3390/rs17010110}, 2025.

\bibitem[{Damiata and Lee(2006)}]{Damiata2006}
Damiata, B.~N. and Lee, T.~C.: Simulated gravitational response to hydraulic testing of unconfined aquifers, Journal of Hydrology, 318, 348--359, \doi{10.1016/j.jhydrol.2005.06.024}, 2006.

\bibitem[{Degen et~al.(2017)Degen, Reinhard, and Cappellaro}]{degen2017}
Degen, C.~L., Reinhard, F., and Cappellaro, P.: Quantum sensing, Reviews of Modern Physics, 89, 035\,002, \doi{10.1103/RevModPhys.89.035002}, 2017.

\bibitem[{Delobbe et~al.(2019)Delobbe, Watlet, Wilfert, and Camp}]{Delobbe2019}
Delobbe, L., Watlet, A., Wilfert, S., and Camp, M.~V.: Exploring the use of underground gravity monitoring to evaluate radar estimates of heavy rainfall, Hydrology and Earth System Sciences, 23, 93--105, \doi{10.5194/hess-23-93-2019}, 2019.

\bibitem[{El-Diasty(2016)}]{El-Diasty2016}
El-Diasty, M.: Groundwater storage change detection using micro-gravimetric technology, \doi{10.1088/1742-2132/13/3/259}, 2016.

\bibitem[{{European Commission}(2025)}]{EC_NEWTONg}
{European Commission}: NEWTON-g: Next Generation Gravity Imager for Field Applications, \urlprefix\url{https://cordis.europa.eu/project/id/101215427}, accessed July 31, 2025, 2025.

\bibitem[{{Exail}(2026)}]{ExailQuantumGravimeters}
{Exail}: Quantum gravimeters, \urlprefix\url{https://www.exail.com/product/quantum-gravimeters}, accessed: 10 April 2026, 2026.

\bibitem[{Faller(1965)}]{Faller1965}
Faller, J.~E.: Results of an absolute determination of the acceleration of gravity, Journal of Geophysical Research, 70, 4035--4038, \doi{10.1029/JZ070i016p04035}, 1965.

\bibitem[{Fang et~al.(2024)}]{Fang2024}
Fang, J. et~al.: Classical and Atomic Gravimetry, To be completed, in press, 2024.

\bibitem[{Farah et~al.(2014)Farah, Guerlin, Landragin, Bouyer, Gaffet, Dos~Santos, and Merlet}]{Farah2014}
Farah, T., Guerlin, C., Landragin, A., Bouyer, P., Gaffet, S., Dos~Santos, F.~P., and Merlet, S.: Underground operation at best sensitivity of the mobile LNE-SYRTE cold atom gravimeter, Gyroscopy and Navigation, 5, 266--274, \doi{10.1134/S2075108714040063}, 2014.

\bibitem[{Forsberg(1984)}]{Forsberg1984}
Forsberg, R.: A Study of Terrain Reductions, Density Anomalies and Geophysical Inversion Methods in Gravity Field Modelling, Tech. rep., Department of Geodetic Science and Surveying, Ohio State University, 1984.

\bibitem[{Francis et~al.(2001)Francis, van Dam, Faller, Amalvict et~al.}]{Francis2001}
Francis, O., van Dam, T., Faller, J.~E., Amalvict, M., et~al.: Results of the International Comparison of Absolute Gravimeters in Walferdange (Luxembourg) of November 1997, Metrologia, 38, 345--348, 2001.

\bibitem[{Freeze and Cherry(1979)}]{Freeze1979}
Freeze, R.~A. and Cherry, J.~A.: Groundwater, vol. 7632, Prentice-Hall Inc., 1979.

\bibitem[{Freier et~al.(2016)Freier, Hauth, Schkolnik, Leykauf, Schilling, Wziontek, Scherneck, M{\"u}ller, and Peters}]{Freier2016GAIN}
Freier, J., Hauth, H., Schkolnik, V., Leykauf, B., Schilling, M., Wziontek, H., Scherneck, H.-G., M{\"u}ller, J., and Peters, A.: Mobile quantum gravity sensor with unprecedented stability, Journal of Physics: Conference Series, 723, 012\,050, \doi{10.1088/1742-6596/723/1/012050}, 2016.

\bibitem[{Gao et~al.(2026)Gao, Wu, Li, and et~al.}]{Gao2026ChipScaleGravimeter}
Gao, L., Wu, W., Li, F., and et~al.: Force balanced chip scale gravimeter achieving record low self noise of 0.1 $\mu$Gal/$\sqrt{\mathrm{Hz}}$, Microsystems \& Nanoengineering, 12, 8, \doi{10.1038/s41378-025-01039-6}, 2026.

\bibitem[{Gehman et~al.(2009)Gehman, Harry, Sanford, Stednick, and Beckman}]{gehman2009}
Gehman, C.~L., Harry, D.~L., Sanford, W.~E., Stednick, J.~D., and Beckman, N.~A.: Estimating specific yield and storage change in an unconfined aquifer using temporal gravity surveys, Water Resources Research, 45, W04\,403, \doi{10.1029/2007WR006096}, 2009.

\bibitem[{Gettings et~al.(2008)Gettings, Chapman, and Allis}]{Gettings2008}
Gettings, P., Chapman, D.~S., and Allis, R.: Techniques, analysis, and noise in a Salt Lake Valley 4D gravity experiment, GEOPHYSICS, 73, \doi{10.1190/1.2996303}, 2008.

\bibitem[{Gillot et~al.(2014)Gillot, Cheng, Imanaliev, Merlet, and Pereira Dos~Santos}]{gillot2016lne_syrte_gravimeter}
Gillot, P., Cheng, B., Imanaliev, A., Merlet, S., and Pereira Dos~Santos, F.: The LNE-SYRTE cold atom gravimeter, Metrologia, \urlprefix\url{https://syrte.obspm.fr/spip/IMG/pdf/pid4177587.pdf}, lNE-SYRTE, Observatoire de Paris, PSL Research University, CNRS, Sorbonne Universit{\'e}s, UPMC Univ. Paris 06, 2014.

\bibitem[{Gonz{\'a}lez~Quir{\'o}s and Fern{\'a}ndez~{\'A}lvarez(2014)}]{gonzalezquiros2014}
Gonz{\'a}lez~Quir{\'o}s, A. and Fern{\'a}ndez~{\'A}lvarez, J.~P.: Simultaneous Solving of Three-Dimensional Gravity Anomalies Caused by Pumping Tests in Unconfined Aquifers, Mathematical Geosciences, 46, 649--664, \doi{10.1007/s11004-014-9539-9}, 2014.

\bibitem[{G{\"u}ntner et~al.(2017)G{\"u}ntner, Reich, Mikolaj, Creutzfeldt, Schroeder, and Wziontek}]{Gntner2017}
G{\"u}ntner, A., Reich, M., Mikolaj, M., Creutzfeldt, B., Schroeder, S., and Wziontek, H.: Landscape-scale water balance monitoring with an iGrav superconducting gravimeter in a field enclosure, Hydrology and Earth System Sciences, 21, 3167--3182, \doi{10.5194/hess-21-3167-2017}, 2017.

\bibitem[{Halloran(2022)}]{Halloran2022}
Halloran, L.~J.: Improving groundwater storage change estimates using time-lapse gravimetry with Gravi4GW, Environmental Modelling and Software, 150, \doi{10.1016/j.envsoft.2022.105340}, 2022.

\bibitem[{Harnisch and Harnisch(2006)}]{Harnisch2006}
Harnisch, G. and Harnisch, M.: Hydrological influences in long gravimetric data series, Journal of Geodynamics, 41, 276--287, \doi{10.1016/j.jog.2005.08.018}, 2006.

\bibitem[{Harrison and Sato(1984)}]{Harrison1984}
Harrison, J.~C. and Sato, T.: Implementation of electrostatic feedback with a LaCoste-Romberg model G gravity meter, Journal of Geophysical Research: Solid Earth, 89, 7957--7961, \doi{10.1029/JB089iB09p07957}, 1984.

\bibitem[{Hasan et~al.(2006)Hasan, Troch, Boll, and Kroner}]{Hasan2006}
Hasan, S., Troch, P.~A., Boll, J., and Kroner, C.: Modeling the Hydrological Effect on Local Gravity at Moxa, Germany, Tech. rep., 2006.

\bibitem[{Hasan et~al.(2008)Hasan, Troch, Bogaart, and Kroner}]{Hasan2008}
Hasan, S., Troch, P.~A., Bogaart, P.~W., and Kroner, C.: Evaluating catchment-scale hydrological modeling by means of terrestrial gravity observations, Water Resources Research, 44, \doi{10.1029/2007WR006321}, 2008.

\bibitem[{Hector et~al.(2013)Hector, {\'S}eguis, Hinderer, Descloitres, Vouillamoz, Wubda, Boy, Luck, and Moigne}]{Hector2013}
Hector, B., {\'S}eguis, L., Hinderer, J., Descloitres, M., Vouillamoz, J.~M., Wubda, M., Boy, J.~P., Luck, B., and Moigne, N.~L.: Gravity effect of water storage changes in a weathered hard-rock aquifer in West Africa: Results from joint absolute gravity, hydrological monitoring and geophysical prospection, Geophysical Journal International, 194, 737--750, \doi{10.1093/gji/ggt146}, 2013.

\bibitem[{Hector et~al.(2015)Hector, S{\'e}guis, Hinderer, Cohard, Wubda, Descloitres, Benarrosh, and Boy}]{Hector2015}
Hector, B., S{\'e}guis, L., Hinderer, J., Cohard, J.~M., Wubda, M., Descloitres, M., Benarrosh, N., and Boy, J.~P.: Water storage changes as a marker for base flow generation processes in a tropical humid basement catchment (Benin): Insights from hybrid gravimetry, Water Resources Research, 51, 8331--8361, \doi{10.1002/2014WR015773}, 2015.

\bibitem[{Heine et~al.(2020)Heine, Matthias, Sahelgozin, Herr, Abend, Timmen, M{\"u}ller, and Rasel}]{Heine2020TransportableQuantumGravimeter}
Heine, N., Matthias, J., Sahelgozin, M., Herr, W., Abend, S., Timmen, L., M{\"u}ller, J., and Rasel, E.~M.: A transportable quantum gravimeter employing delta-kick collimated Bose--Einstein condensates, European Physical Journal D, 74, 174, \doi{10.1140/epjd/e2020-10120-x}, 2020.

\bibitem[{Heiskanen and Moritz(1967)}]{HeiskanenMoritz1967}
Heiskanen, W.~A. and Moritz, H.: Physical Geodesy, W. H. Freeman and Company, San Francisco, 1967.

\bibitem[{Herckenrath et~al.(2012)Herckenrath, Auken, Christiansen, Behroozmand, and Bauer-Gottwein}]{herckenrath2012}
Herckenrath, D., Auken, E., Christiansen, L., Behroozmand, A.~A., and Bauer-Gottwein, P.: Coupled hydrogeophysical inversion using time-lapse magnetic resonance sounding and time-lapse gravity data for hydraulic aquifer testing: Will it work in practice?, Water Resources Research, 48, W01\,539, \doi{10.1029/2011WR010411}, 2012.

\bibitem[{Hern{\'a}ndez-S{\'a}nchez et~al.(2021)Hern{\'a}ndez-S{\'a}nchez, Castellanos, Herrera-Barrientos, and Belmonte-Jim{\'e}nez}]{Hernndez-Snchez2021}
Hern{\'a}ndez-S{\'a}nchez, R.~I., Castellanos, F., Herrera-Barrientos, J., and Belmonte-Jim{\'e}nez, S.~I.: Gravimetric Determination of Storage Coefficient and Storage Change of Groundwater in an Uncontrolled and Unconfined Aquifer, Natural Resources Research, 30, 4207--4218, \doi{10.1007/s11053-021-09904-7}, 2021.

\bibitem[{Hinderer et~al.(2006)Hinderer, Andersen, Lemoine, Crossley, and Boy}]{Hinderer2006}
Hinderer, J., Andersen, O., Lemoine, F., Crossley, D., and Boy, J.~P.: Seasonal changes in the European gravity field from GRACE: A comparison with superconducting gravimeters and hydrology model predictions, Journal of Geodynamics, 41, 59--68, \doi{10.1016/j.jog.2005.08.037}, 2006.

\bibitem[{Hinderer et~al.(2007)Hinderer, Crossley, and Warburton}]{Hinderer2007}
Hinderer, J., Crossley, D., and Warburton, R.~J.: Superconducting gravimetry, Treatise on Geophysics, 3, 65--122, 2007.

\bibitem[{Hinnell et~al.(2010)Hinnell, Ferr{\'e}, Vrugt, Huisman, Moysey, Rings, and Kowalsky}]{Hinnell2010}
Hinnell, A.~C., Ferr{\'e}, T. P.~A., Vrugt, J.~A., Huisman, J.~A., Moysey, S., Rings, J., and Kowalsky, M.~B.: Improved extraction of hydrologic information from geophysical data through coupled hydrogeophysical inversion, Water Resources Research, 46, W00D40, \doi{10.1029/2008WR007060}, 2010.

\bibitem[{Hokkanen et~al.(2006)Hokkanen, Korhonen, and Virtanen}]{Hokkanen2006}
Hokkanen, T., Korhonen, K., and Virtanen, H.: Hydrogeological Effects on Superconducting Gravimeter Measurements at Mets{\"a}hovi in Finland, Tech. rep., \urlprefix\url{http://library.seg.org/}, 2006.

\bibitem[{Hokkanen et~al.(2007)Hokkanen, Korhonen, Virtanen, and Laine}]{Hokkanen2007}
Hokkanen, T., Korhonen, K., Virtanen, H., and Laine, E.~L.: Effects of the fracture water of bedrock on superconducting gravimeter data, Near Surface Geophysics, 5, 133--139, \doi{10.3997/1873-0604.2006025}, 2007.

\bibitem[{Hsiao et~al.(2021)Hsiao, Chang, Yang, and Tseng}]{Hsiao2021}
Hsiao, Y.~S., Chang, J.~C., Yang, R.~J., and Tseng, T.~P.: Estimating the specific yield in an unconfined aquifer using the gravimetric method: a case study in the Zhoushui River alluvial fan, Journal of the Chinese Institute of Engineers, Transactions of the Chinese Institute of Engineers,Series A, 44, 820--830, \doi{10.1080/02533839.2021.1978328}, 2021.

\bibitem[{Imanishi et~al.(2013)Imanishi, Nawa, and Takayama}]{Imanishi2013}
Imanishi, Y., Nawa, K., and Takayama, H.: Local hydrological processes in a fractured bedrock and the short-term effect on gravity at Matsushiro, Japan, Journal of Geodynamics, 63, 62--68, \doi{10.1016/j.jog.2012.10.001}, 2013.

\bibitem[{Jakeman et~al.(2016)Jakeman, Barreteau, Hunt, Rinaudo, and Ross}]{jakeman2016}
Jakeman, A.~J., Barreteau, O., Hunt, R.~J., Rinaudo, J.-D., and Ross, A.: Integrated Groundwater Management: Concepts, Approaches and Challenges, Springer, \doi{10.1007/978-3-319-23576-9}, 2016.

\bibitem[{Kaban et~al.(2015)Kaban, Mooney, and Petrunin}]{kaban2015cratonic}
Kaban, M.~K., Mooney, W.~D., and Petrunin, A.~G.: Cratonic root beneath North America shifted by basal drag from the convecting mantle, Nature Geoscience, 8, 797--800, \doi{10.1038/ngeo2525}, 2015.

\bibitem[{Kaban et~al.(2016)Kaban, Stolk, Tesauro, El~Khrepy, Al-Arifi, Beekman, and Cloetingh}]{kaban2016}
Kaban, M.~K., Stolk, W., Tesauro, M., El~Khrepy, S., Al-Arifi, N., Beekman, F., and Cloetingh, S. A. P.~L.: 3D density model of the upper mantle of Asia based on inversion of gravity and seismic tomography data, Geochemistry, Geophysics, Geosystems, 17, 4457--4477, \doi{https://doi.org/10.1002/2016GC006458}, 2016.

\bibitem[{Karcher et~al.(2018)Karcher, Imanaliev, Merlet et~al.}]{Karcher2018UltracoldSourcesInterferometers}
Karcher, R., Imanaliev, A., Merlet, S., et~al.: Improving the accuracy of atom interferometers with ultracold sources, New Journal of Physics, 20, 113\,041, \doi{10.1088/1367-2630/aaf07d}, 2018.

\bibitem[{Karl(2005)}]{KARL2005183}
Karl, W.: 3.6 - Regularization in Image Restoration and Reconstruction, in: Handbook of Image and Video Processing (Second Edition), edited by BOVIK, A., Communications, Networking and Multimedia, pp. 183--V, Academic Press, Burlington, second edition edn., ISBN 978-0-12-119792-6, \doi{https://doi.org/10.1016/B978-012119792-6/50075-9}, 2005.

\bibitem[{Kasevich and Chu(1991)}]{Kasevich1991}
Kasevich, M. and Chu, S.: Atomic interferometry using stimulated Raman transitions, Physical Review Letters, 67, 181--184, \doi{10.1103/PhysRevLett.67.181}, 1991.

\bibitem[{Kennedy et~al.(2016)Kennedy, Ferre, and Creutzfeldt}]{Kennedy2016}
Kennedy, J., Ferre, T.~P., and Creutzfeldt, B.: Time-lapse gravity data for monitoring and modeling artificial recharge through a thick unsaturated zone, Water Resources Research, 52, 7244--7261, \doi{10.1002/2016WR018770}, 2016.

\bibitem[{Kennedy and Larsen(2023)}]{Kennedy2023}
Kennedy, J.~R. and Larsen, J.~D.: Heavy: Software for forward modeling gravity change from MODFLOW output, Environmental Modelling and Software, 165, \doi{10.1016/j.envsoft.2023.105714}, 2023.

\bibitem[{Kennedy et~al.(2022)Kennedy, Wildermuth, Knight, and Larsen}]{Kennedy2022}
Kennedy, J.~R., Wildermuth, L., Knight, J.~E., and Larsen, J.: Improving Groundwater Model Calibration with Repeat Microgravity Measurements, Groundwater, 60, 393--403, \doi{10.1111/gwat.13167}, 2022.

\bibitem[{Krause et~al.(2009)Krause, Naujoks, Fink, and Kroner}]{Krause2009}
Krause, P., Naujoks, M., Fink, M., and Kroner, C.: The impact of soil moisture changes on gravity residuals obtained with a superconducting gravimeter, Journal of Hydrology, 373, 151--163, \doi{10.1016/j.jhydrol.2009.04.019}, 2009.

\bibitem[{Kroner and Jahr(2006)}]{Kroner2006}
Kroner, C. and Jahr, T.: Hydrological experiments around the superconducting gravimeter at Moxa Observatory, Journal of Geodynamics, 41, 268--275, \doi{10.1016/j.jog.2005.08.012}, 2006.

\bibitem[{Kroner et~al.(2009)Kroner, Thomas, Dobslaw, Abe, and Weise}]{Kroner2009}
Kroner, C., Thomas, M., Dobslaw, H., Abe, M., and Weise, A.: Seasonal effects of non-tidal oceanic mass shifts in observations with superconducting gravimeters, Journal of Geodynamics, 48, 354--359, \doi{10.1016/j.jog.2009.09.009}, 2009.

\bibitem[{Kumar et~al.(2022)Kumar, {Augusto de Jesus Pacheco}, Kaushik, and Rodrigues}]{KUMAR2022}
Kumar, A., {Augusto de Jesus Pacheco}, D., Kaushik, K., and Rodrigues, J.~J.: Futuristic view of the Internet of Quantum Drones: Review, challenges and research agenda, Vehicular Communications, 36, 100\,487, \doi{https://doi.org/10.1016/j.vehcom.2022.100487}, 2022.

\bibitem[{Lachassagne et~al.(2021)Lachassagne, Dewandel, and Wyns}]{Lachassagne2021}
Lachassagne, P., Dewandel, B., and Wyns, R.: Review: Hydrogeology of weathered crystalline/hard-rock aquifers---guidelines for the operational survey and management of their groundwater resources, Hydrogeology Journal, 29, 2561--2594, \doi{10.1007/s10040-021-02339-7}, 2021.

\bibitem[{LaCoste(1934)}]{LaCoste1934}
LaCoste, L. J.~B.: A New Type Long Period Vertical Seismograph, Physics, 5, 178--180, \doi{10.1063/1.1745243}, 1934.

\bibitem[{Lampitelli and Francis(2010)}]{Lampitelli2010}
Lampitelli, C. and Francis, O.: Hydrological effects on gravity and correlations between gravitational variations and level of the Alzette River at the station of Walferdange, Luxembourg, Journal of Geodynamics, 49, 31--38, \doi{10.1016/j.jog.2009.08.003}, 2010.

\bibitem[{Leiriao et~al.(2009)Leiriao, He, Christiansen, Andersen, and Bauer-Gottwein}]{Leirio2009}
Leiriao, S., He, X., Christiansen, L., Andersen, O.~B., and Bauer-Gottwein, P.: Calculation of the temporal gravity variation from spatially variable water storage change in soils and aquifers, Journal of Hydrology, 365, 302--309, \doi{10.1016/j.jhydrol.2008.11.040}, - calculation of gravity response from hydrogeological model<br/>-prism formula, the MacMillan formula and the point-mass approximation, 2009.

\bibitem[{Lewis and Rodgers(2018)}]{LewisRodgers2018}
Lewis, R. and Rodgers, D.: Preparatory Tests 2016--17, Tech. Rep. JRC110440, Joint Research Centre (JRC), \urlprefix\url{https://publications.jrc.ec.europa.eu/repository/bitstream/JRC110440/jrc110440_technical_note_preparatory_tests_2016-17-libre.pdf}, 2018.

\bibitem[{Longuevergne et~al.(2009)Longuevergne, Boy, Florsch, Viville, Ferhat, Ulrich, Luck, and Hinderer}]{Longuevergne2009}
Longuevergne, L., Boy, J.~P., Florsch, N., Viville, D., Ferhat, G., Ulrich, P., Luck, B., and Hinderer, J.: Local and global hydrological contributions to gravity variations observed in Strasbourg, Journal of Geodynamics, 48, 189--194, \doi{10.1016/j.jog.2009.09.008}, 2009.

\bibitem[{Luan et~al.(2023)Luan, Shen, and Jia}]{Luan2023}
Luan, W., Shen, W., and Jia, J.: Analysis of iGrav Superconducting Gravity Measurements in Kunming, China, with Emphasis on Calibration, Tides, and Hydrology, Pure and Applied Geophysics, 180, 643--660, \doi{10.1007/s00024-022-03036-6}, 2023.

\bibitem[{MacMillan(1958)}]{MacMillan1958}
MacMillan, W. D. W.~D.: The Theory of the Potential, Dover Publications, bibliography: p. 463-465, 1958.

\bibitem[{Maina et~al.(2021)Maina, Guadagnini, and Riva}]{Maina2021}
Maina, F.~Z., Guadagnini, A., and Riva, M.: Impact of multiple uncertainties on gravimetric variations across randomly heterogeneous aquifers during pumping, Advances in Water Resources, 154, \doi{10.1016/j.advwatres.2021.103978}, 2021.

\bibitem[{Maloszewski and Zuber(1985)}]{Maloszewski1985}
Maloszewski, P. and Zuber, A.: On the Theory of Tracer Experiments in Fissured Rocks with a Porous Matrix, Tech. rep., 1985.

\bibitem[{Mikolaj et~al.(2015)Mikolaj, Meurers, and Mojze{\v s}}]{Mikolaj2015}
Mikolaj, M., Meurers, B., and Mojze{\v s}, M.: The reduction of hydrology-induced gravity variations at sites with insufficient hydrological instrumentation, Studia Geophysica et Geodaetica, 59, 424--437, \doi{10.1007/s11200-014-0232-8}, 2015.

\bibitem[{Mikolaj et~al.(2019)Mikolaj, Reich, and Guentner}]{Mikolaj2019}
Mikolaj, M., Reich, M., and Guentner, A.: Resolving Geophysical Signals by Terrestrial Gravimetry: A Time Domain Assessment of the Correction-Induced Uncertainty, Journal of Geophysical Research: Solid Earth, 124, 2487--2504, \doi{10.1029/2018JB016682}, 2019.

\bibitem[{Mohammadi et~al.(2026)Mohammadi, Mohammadigheymasi, and Halloran}]{Mohammadi2026GravCHAW}
Mohammadi, N., Mohammadigheymasi, H., and Halloran, L. J.~S.: GravCHAW: A Software Framework for the Assimilation of Time-Lapse Gravimetry Data in Groundwater Models, Computers \& Geosciences, 209, 106\,118, \doi{10.1016/j.cageo.2026.106118}, 2026.

\bibitem[{Murty and Raghavan(2002)}]{Murty2002}
Murty, B.~V. and Raghavan, V.~K.: The gravity method in groundwater exploration in crystalline rocks: A study in the peninsular granitic region of Hyderabad, India, Hydrogeology Journal, 10, 307--321, \doi{10.1007/s10040-001-0184-2}, 2002.

\bibitem[{M{\'e}noret et~al.(2018)M{\'e}noret, Vermeulen, Le~Moigne, Bonvalot, Bouyer, Landragin, and Desruelle}]{Menoret2018}
M{\'e}noret, V., Vermeulen, P., Le~Moigne, N., Bonvalot, S., Bouyer, P., Landragin, A., and Desruelle, B.: Gravity measurements below $10^{-9}$ g with a transportable absolute quantum gravimeter, Scientific Reports, 8, 12\,300, \doi{10.1038/s41598-018-30608-1}, 2018.

\bibitem[{Nagy(1966)}]{Nagy1966}
Nagy, D.: The Gravitational Attraction of a Right Rectangular Prism, Tech. rep., \urlprefix\url{http://library.seg.org/}, 1966.

\bibitem[{Nagy et~al.(2000)Nagy, Papp, and Benedek}]{Nagy2000}
Nagy, D., Papp, G., and Benedek, J.: The gravitational potential and its derivatives for the prism, Journal of Geodesy, 74, 552--560, \doi{10.1007/s001900000116}, 2000.

\bibitem[{Narducci et~al.(2022)Narducci, Black, and Burke}]{narducci2022}
Narducci, F.~A., Black, A.~T., and Burke, J.~H.: Advances toward fieldable atom interferometers, Advances in Physics: X, 7, \doi{10.1080/23746149.2021.1946426}, 2022.

\bibitem[{Naujoks et~al.(2010)Naujoks, Kroner, Weise, Jahr, Krause, and Eisner}]{Naujoks2010}
Naujoks, M., Kroner, C., Weise, A., Jahr, T., Krause, P., and Eisner, S.: Evaluating local hydrological modelling by temporal gravity observations and a gravimetric three-dimensional model, Geophysical Journal International, 182, 233--249, \doi{10.1111/j.1365-246X.2010.04615.x}, 2010.

\bibitem[{Neumeyer et~al.(2006)Neumeyer, Barthelmes, Dierks, Flechtner, Harnisch, Harnisch, Hinderer, Imanishi, Kroner, Meurers, Petrovic, Reigber, Schmidt, Schwintzer, Sun, and Virtanen}]{Neumeyer2006}
Neumeyer, J., Barthelmes, F., Dierks, O., Flechtner, F., Harnisch, M., Harnisch, G., Hinderer, J., Imanishi, Y., Kroner, C., Meurers, B., Petrovic, S., Reigber, C., Schmidt, R., Schwintzer, P., Sun, H.~P., and Virtanen, H.: Combination of temporal gravity variations resulting from superconducting gravimeter (SG) recordings, GRACE satellite observations and global hydrology models, Journal of Geodesy, 79, 573--585, \doi{10.1007/s00190-005-0014-8}, 2006.

\bibitem[{Neumeyer et~al.(2008)Neumeyer, Barthelmes, Kroner, Petrovic, Schmidt, Virtanen, and Wilmes}]{Neumeyer2008}
Neumeyer, J., Barthelmes, F., Kroner, C., Petrovic, S., Schmidt, R., Virtanen, H., and Wilmes, H.: Analysis of gravity field variations derived from Superconducting Gravimeter recordings, the GRACE satellite and hydrological models at selected European sites, Tech. rep., \urlprefix\url{http://isdc.gfz-potsdam.de/grace}, 2008.

\bibitem[{Niebauer et~al.(1995)Niebauer, Sasagawa, Faller, Hilt, and Klopping}]{Niebauer1995}
Niebauer, T.~M., Sasagawa, G.~S., Faller, J.~E., Hilt, R., and Klopping, F.: A new generation of absolute gravimeters, Metrologia, 32, 159, \doi{10.1088/0026-1394/32/3/004}, 1995.

\bibitem[{Oh et~al.(2024)Oh, Gregoire, Black, Hughes, Kunz, Larsen, Lautier-Gaud, Lee, Schwindt, Mouradian, Narducci, and Sackett}]{Oh2024}
Oh, E., Gregoire, M.~D., Black, A.~T., Hughes, K.~J., Kunz, P.~D., Larsen, M., Lautier-Gaud, J., Lee, J., Schwindt, P. D.~D., Mouradian, S.~L., Narducci, F.~A., and Sackett, C.~A.: A Perspective on Quantum Sensors from Basic Research to Commercial Applications, arXiv preprint arXiv:2407.00689, pp. 1--96, \doi{10.48550/arXiv.2407.00689}, 2024.

\bibitem[{O'Neill et~al.(2026)O'Neill, Rodell, and Loomis}]{ONeill2026}
O'Neill, M.~M., Rodell, M., and Loomis, B.: Spatially Refined Global Terrestrial Water Storage Trends and Annual Cycles from GRACE and GRACE-FO, in: EGU General Assembly 2026, pp. EGU26--16\,019, Vienna, Austria, \doi{10.5194/egusphere-egu26-16019}, 2026.

\bibitem[{Pendiuk et~al.(2020)Pendiuk, Guarracino, Reich, Brunini, and G{\"u}ntner}]{Pendiuk2020}
Pendiuk, J.~E., Guarracino, L., Reich, M., Brunini, C., and G{\"u}ntner, A.: Estimating the specific yield of the Pampeano aquifer, Argentina, using superconducting gravimeter data, Hydrogeology Journal, 28, 2303--2313, \doi{10.1007/s10040-020-02212-z}, 2020.

\bibitem[{Peters et~al.(2001)Peters, Chung, and Chu}]{Peters2001}
Peters, A., Chung, K.~Y., and Chu, S.: High-precision gravity measurements using atom interferometry, Metrologia, 38, 25, \doi{10.1088/0026-1394/38/1/4}, 2001.

\bibitem[{Petrovi{\'c}(1996)}]{Petrovic1996}
Petrovi{\'c}, S.: Determination of the Potential of Homogeneous Polyhedral Bodies Using Line Integrals, Journal of Geodesy, 71, 44--52, \doi{10.1007/s001900050074}, 1996.

\bibitem[{Piccolroaz et~al.(2015)Piccolroaz, Majone, Palmieri, Cassiani, and Bellin}]{Piccolroaz2015}
Piccolroaz, S., Majone, B., Palmieri, F., Cassiani, G., and Bellin, A.: On the use of spatially distributed, time-lapse microgravity surveys to inform hydrological modeling, Water Resources Research, 51, 7280--7298, \doi{10.1002/2015WR016994}, 2015.

\bibitem[{Pivetta et~al.(2021)Pivetta, Braitenberg, Gabrov{\v s}ek, Gabriel, and Meurers}]{Pivetta2021}
Pivetta, T., Braitenberg, C., Gabrov{\v s}ek, F., Gabriel, G., and Meurers, B.: Gravity as a tool to improve the hydrologic mass budget in karstic areas, Hydrology and Earth System Sciences, 25, 6001--6021, \doi{10.5194/hess-25-6001-2021}, 2021.

\bibitem[{Pool(2008)}]{Pool2008}
Pool, D.~R.: The utility of gravity and water-level monitoring at alluvial aquifer wells in southern Arizona, GEOPHYSICS, 73, \doi{10.1190/1.2980395}, 2008.

\bibitem[{Pool and Eychaner(1995)}]{Pool1995}
Pool, D.~R. and Eychaner, J.~H.: Measurements of Aquifer-Storage Change and Specific Yield Using Gravity Surveys, Groundwater, 33, 425--432, \doi{10.1111/j.1745-6584.1995.tb00299.x}, 1995.

\bibitem[{Prasad et~al.(2022)Prasad, Middlemiss, Noack et~al.}]{prasad2022}
Prasad, A., Middlemiss, R.~P., Noack, A., et~al.: A 19 day earth tide measurement with a MEMS gravimeter, Scientific Reports, 12, 13\,091, \doi{10.1038/s41598-022-16881-1}, 2022.

\bibitem[{Prothero and Goodkind(1968)}]{Prothero1968}
Prothero, W.~A. and Goodkind, J.~M.: A Superconducting Gravimeter, Review of Scientific Instruments, 39, 1257--1262, \doi{10.1063/1.1683651}, 1968.

\bibitem[{Ramillien et~al.(2008)Ramillien, Famiglietti, and Wahr}]{Ramillien2008}
Ramillien, G., Famiglietti, J.~S., and Wahr, J.: Detection of continental hydrology and glaciology Signals from GRACE: A review, \doi{10.1007/s10712-008-9048-9}, 2008.

\bibitem[{Rau et~al.(2020)Rau, Cuthbert, Post, Andersen, Befus, Devlin, Kurylyk, Li, Morgan, Rowland, Shanafield, Simmons, Ward, and Zlotnik}]{rau2020}
Rau, G.~C., Cuthbert, M.~O., Post, V. E.~A., Andersen, M.~S., Befus, K.~M., Devlin, J.~F., Kurylyk, B.~L., Li, L., Morgan, L.~K., Rowland, J.~C., Shanafield, M.~A., Simmons, C.~T., Ward, A.~S., and Zlotnik, V.~A.: Future-proofing hydrogeology by revising groundwater monitoring practice, Hydrogeology Journal, 28, 2963--2969, \doi{10.1007/s10040-020-02242-7}, 2020.

\bibitem[{Reich et~al.(2021)Reich, Mikolaj, Blume, and G{\"u}ntner}]{Reich2021}
Reich, M., Mikolaj, M., Blume, T., and G{\"u}ntner, A.: Field-Scale Subsurface Flow Processes Inferred From Continuous Gravity Monitoring During a Sprinkling Experiment, Water Resources Research, 57, \doi{10.1029/2021WR030044}, 2021.

\bibitem[{Rodell and Reager(2023)}]{Rodell2023}
Rodell, M. and Reager, J.~T.: Water cycle science enabled by the GRACE and GRACE-FO satellite missions, Nature Water, 1, 47--59, \doi{10.1038/s44221-022-00005-0}, 2023.

\bibitem[{Sasagawa et~al.(1989)Sasagawa, Zumberge, Mark~Stevenson, Parker, Ander, and Gorman}]{Sasagawa1989}
Sasagawa, G.~S., Zumberge, M.~A., Mark~Stevenson, J., Parker, R.~L., Ander, M.~E., and Gorman, R.~W.: The 1987 southeastern Alaska gravity calibration range: Absolute and relative gravity measurements, Journal of Geophysical Research: Solid Earth, 94, 7661--7665, \doi{10.1029/JB094iB06p07661}, 1989.

\bibitem[{Scanlon et~al.(2023)Scanlon, Fakhreddine, Rateb, de~Graaf, Famiglietti, Gleeson, Grafton, Jobbagy, Kebede, Kolusu, Konikow, Long, Mekonnen, Schmied, Mukherjee, MacDonald, Reedy, Shamsudduha, Simmons, Sun, Taylor, Villholth, V{\"o}r{\"o}smarty, and Zheng}]{Scanlon2023}
Scanlon, B.~R., Fakhreddine, S., Rateb, A., de~Graaf, I., Famiglietti, J., Gleeson, T., Grafton, R.~Q., Jobbagy, E., Kebede, S., Kolusu, S.~R., Konikow, L.~F., Long, D., Mekonnen, M., Schmied, H.~M., Mukherjee, A., MacDonald, A., Reedy, R.~C., Shamsudduha, M., Simmons, C.~T., Sun, A., Taylor, R.~G., Villholth, K.~G., V{\"o}r{\"o}smarty, C.~J., and Zheng, C.: Global water resources and the role of groundwater in a resilient water future, Nature Reviews Earth and Environment, 4, 87--101, \doi{10.1038/s43017-022-00378-6}, 2023.

\bibitem[{Schilling et~al.(2020)Schilling, Wodey, Timmen, Tell, Zipfel, Schlippert, Schubert, Rasel, and M{\"u}ller}]{Schilling2020VLBAIGrav}
Schilling, M., Wodey, {\'E}., Timmen, L., Tell, D., Zipfel, K.~H., Schlippert, D., Schubert, C., Rasel, E.~M., and M{\"u}ller, J.: Gravity field modelling for the Hannover 10 m atom interferometer, Journal of Geodesy, 94, 122, \doi{10.1007/s00190-020-01451-y}, 2020.

\bibitem[{Schroeter et~al.(2025)Schroeter, Orme, Lehmann, Lehmann, Chaudhari, K{\"u}sel, Wang, Hildebrandt, Totsche, Trumbore, and Gleixner}]{Schroeter2025}
Schroeter, S.~A., Orme, A.~M., Lehmann, K., Lehmann, R., Chaudhari, N.~M., K{\"u}sel, K., Wang, H., Hildebrandt, A., Totsche, K.~U., Trumbore, S., and Gleixner, G.: Hydroclimatic extremes threaten groundwater quality and stability, Nature communications, 16, 720, \doi{10.1038/s41467-025-55890-2}, 2025.

\bibitem[{Seraphin et~al.(2018)Seraphin, Gon{\c c}alv{\`e}s, Vallet-Coulomb et~al.}]{seraphin2018}
Seraphin, P., Gon{\c c}alv{\`e}s, J., Vallet-Coulomb, C., et~al.: Multi-approach assessment of the spatial distribution of the specific yield: application to the Crau plain aquifer, France, Hydrogeology Journal, 26, 1221--1238, \doi{10.1007/s10040-018-1753-y}, 2018.

\bibitem[{Sharp(2023)}]{SharpJr2023}
Sharp, J.: A Glossary of Hydrogeology, The Groundwater Project, \doi{10.21083/978-1-77470-079-2}, 2023.

\bibitem[{Shettell et~al.(2024)Shettell, Lee, Oon, Maksimova, Hufnagel, Wei, and Dumke}]{Shettell2024}
Shettell, N., Lee, K.~S., Oon, F.~E., Maksimova, E., Hufnagel, C., Wei, S., and Dumke, R.: Geophysical survey based on hybrid gravimetry using relative measurements and an atomic gravimeter as an absolute reference, Scientific Reports, 14, \doi{10.1038/s41598-024-61447-0}, 2024.

\bibitem[{Telford et~al.(1990)Telford, Geldart, and Sheriff}]{telford1990applied}
Telford, W.~M., Geldart, L.~P., and Sheriff, R.~E.: Applied Geophysics, Cambridge University Press, Cambridge, 2 edn., 1990.

\bibitem[{Tikhonov and Arsenin(1977)}]{tikhonov1977illposed}
Tikhonov, A.~N. and Arsenin, V.~Y.: Solutions of Ill-Posed Problems, Scripta Series in Mathematics, V. H. Winston \& Sons, Washington, DC, ISBN 9780470991244, translated from Russian by Fritz John, 1977.

\bibitem[{Torge et~al.(2025)Torge, M{\"u}ller, and Pail}]{torge2025}
Torge, W., M{\"u}ller, J., and Pail, R.: Geodesy, De Gruyter, Berlin, Germany, 5 edn., ISBN 978-3-11-044201-9, 2025.

\bibitem[{UNESCO(2022)}]{UNESCO2022}
UNESCO: Groundwater : making the invisible visible, p. 225, 2022.

\bibitem[{Van~Camp et~al.(2017)Van~Camp, de~Viron, Watlet, Meurers, Francis, and Caudron}]{VanCamp2017}
Van~Camp, M., de~Viron, O., Watlet, A., Meurers, B., Francis, O., and Caudron, C.: Geophysics from terrestrial time-variable gravity measurements, Reviews of Geophysics, 55, 938--992, \doi{10.1002/2017RG000566}, 2017.

\bibitem[{Veryaskin et~al.(2022)}]{Veryaskin2022}
Veryaskin, A.~V. et~al.: Mobile atom interferometry and its applications in geophysics, arXiv preprint, \urlprefix\url{https://arxiv.org/abs/2108.05519}, 2022.

\bibitem[{Wagner et~al.(2025)Wagner, Duethmann, Kiesel, Pool, Hrachowitz, Ceola, Herzog, Houska, Loritz, Spieler, Staudinger, Tarasova, Thober, Fohrer, Tetzlaff, Wagener, and Guse}]{Wagner2025}
Wagner, P.~D., Duethmann, D., Kiesel, J., Pool, S., Hrachowitz, M., Ceola, S., Herzog, A., Houska, T., Loritz, R., Spieler, D., Staudinger, M., Tarasova, L., Thober, S., Fohrer, N., Tetzlaff, D., Wagener, T., and Guse, B.: The Unexploited Treasures of Hydrological Observations Beyond Streamflow for Catchment Modeling, WIREs Water, 12, e70\,018, \doi{https://doi.org/10.1002/wat2.70018}, e70018 WATER-972.R2, 2025.

\bibitem[{Warburton et~al.(2010)Warburton, Peirce, Meyer, and Goodkind}]{Warburton2010}
Warburton, R.~J., Peirce, J., Meyer, C., and Goodkind, J.~M.: Initial results with the new GWR iGrav\texttrademark{} superconducting gravity meter, Society of Exploration Geophysicists (SEG) Technical Program Expanded Abstracts, pp. 1141--1145, \doi{10.1190/1.3513028}, 2010.

\bibitem[{Weise and Jahr(2018)}]{Weise2018}
Weise, A. and Jahr, T.: The Improved Hydrological Gravity Model for Moxa Observatory, Germany, Pure and Applied Geophysics, 175, 1755--1763, \doi{10.1007/s00024-017-1546-6}, 2018.

\bibitem[{Weise et~al.(2009)Weise, Kroner, Abe, Ihde, Jentzsch, Naujoks, Wilmes, and Wziontek}]{Weise2009}
Weise, A., Kroner, C., Abe, M., Ihde, J., Jentzsch, G., Naujoks, M., Wilmes, H., and Wziontek, H.: Gravity field variations from superconducting gravimeters for GRACE validation, Journal of Geodynamics, 48, 325--330, \doi{10.1016/j.jog.2009.09.034}, 2009.

\bibitem[{Wu et~al.(2019)Wu, Pagel, Malek, Nguyen, Zi, Scheirer, and M{\"u}ller}]{Wu2019}
Wu, X.~J., Pagel, Z., Malek, B.~S., Nguyen, T.~H., Zi, F., Scheirer, D.~S., and M{\"u}ller, H.: Gravity surveys using a mobile atom interferometer, Science Advances, 5, eaax0800, \doi{10.1126/sciadv.aax0800}, 2019.

\bibitem[{Wziontek et~al.(2009)Wziontek, Wilmes, Wolf, Werth, and G{\"u}ntner}]{Wziontek2009}
Wziontek, H., Wilmes, H., Wolf, P., Werth, S., and G{\"u}ntner, A.: Time series of superconducting gravimeters and water storage variations from the global hydrology model WGHM, Journal of Geodynamics, 48, 166--171, \doi{10.1016/j.jog.2009.09.036}, 2009.

\bibitem[{Zhong et~al.(2022)Zhong, Ren, Tang, Lin, Chen, Deng, and Jiang}]{Zhong2022ConstrainedGravity}
Zhong, Y., Ren, Z., Tang, J., Lin, Y., Chen, B., Deng, Y., and Jiang, Y.: Constrained Gravity Inversion With Adaptive Inversion Grid Refinement in Spherical Coordinates and Its Application to Mantle Structure Beneath Tibetan Plateau, Journal of Geophysical Research: Solid Earth, 127, e2021JB022\,916, \doi{10.1029/2021JB022916}, 2022.

\end{thebibliography}

\end{document}